\documentclass[
aps,                    
prd,                    
superscriptaddress,    
nofootinbib,            
amsmath,               %
amssymb,               %
a4paper,                
floatfix]               
{revtex4}               

\usepackage{booktabs}
\usepackage{multirow}
\usepackage[colorlinks = true,
            linkcolor = blue,
            urlcolor  = blue,
            citecolor = blue,
            anchorcolor = blue]{hyperref}

\usepackage{float}
\usepackage{amssymb}
\usepackage{graphics}
\usepackage{graphicx}
\usepackage{amsmath}
\usepackage{hyperref}
\usepackage{amsfonts}
\usepackage{subfigure}
\usepackage{xcolor}
\usepackage{cancel}
\usepackage{colordvi}

\newcommand{\fr}{\frac}

\newcommand{\la}{\lambda}
\newcommand{\La}{\Lambda}
\newcommand{\si}{\sigma}

\newcommand{\ga}{\gamma}

\newcommand{\de}{\delta}

\newcommand{\be}{\begin{equation}}
\newcommand{\ee}{\end{equation}}
\newcommand{\beqa}{\begin{eqnarray}}
\newcommand{\eeqa}{\end{eqnarray}}
\newcommand{\bi}{\begin{itemize}}
\newcommand{\ei}{\end{itemize}}
\newcommand{\ben}{\begin{enumerate}}
\newcommand{\een}{\end{enumerate}}

\begin{document}

\title{The Ages of the Oldest Astrophysical Objects in an Ellipsoidal Universe}

\author{Selinay Sude Binici}
\email{binici.sude@gmail.com}
\affiliation{Department of Physics, Mimar Sinan Fine Arts University, Bomonti 34380, \.{I}stanbul, T\"urkiye}

\author{Cemsinan Deliduman}
\email{cdeliduman@gmail.com}
\affiliation{Department of Physics, Mimar Sinan Fine Arts University, Bomonti 34380, \.{I}stanbul, T\"urkiye}

\author{Furkan \c{S}akir Dilsiz}
\email[Corresponding author: ]{furkansakirdilsiz@gmail.com}
\affiliation{Department of Physics, Mimar Sinan Fine Arts University, Bomonti 34380, \.{I}stanbul, T\"urkiye}

\begin{abstract}
James Webb Space Telescope's (JWST) observations since its launch have shown us that there could be very massive and very large galaxies, as well as massive quasars very early in the history of the Universe, conflicting expectations of the $\Lambda$CDM model. This so-called ``impossibly early galaxy problem'' requires too rapid star formation in the earliest galaxies than appears to be permitted by the $\Lambda$CDM model. In fact, this might not be a high masses problem, but a ``time-compression problem'': time too short for the observed large and massive structures to form from the initial seeds. A cosmological model that could allocate more time for the earliest large structures to form would be more conforming to the data than the $\Lambda$CDM model. In this work we are going to discuss how the recently proposed $\gamma\delta$CDM model might ease and perhaps resolve the time-compression problem. In the $\gamma\delta$CDM model, different energy densities contribute to the Hubble parameter with different weights. Additionally, in the formula for the Hubble parameter, energy densities depend on the redshift differently than what their physical nature dictates. This new way of relating Universe's energy content to the Hubble parameter leads to a modified relation between cosmic time and redshift. We test the observational relevance of the $\gamma\delta$CDM model to the age problem by constraining its parameters with the ages of the oldest astronomical objects (OAO) together with the cosmic chronometers (CC) Hubble data and the Pantheon+ Type Ia supernovae data of the late Universe at low redshift. We find that, thanks to a modified time-redshift relation, the $\gamma\delta$CDM model has a more plausible time period at high redshift for large and massive galaxies and massive quasars to form, whereas the age of the Universe today is not modified significantly.
\end{abstract}

\maketitle


\section{Introduction}

Since the seminal paper of Edwin Hubble which declared an expanding Universe instead of a static one \cite{Hubble:1929ig}, the degree of expansion, which is given by the value of the Hubble constant $H_0$, has been subject of controversy and debate among the observational and theoretical cosmologists. Hubble's original calculation gave a too large value of $H_0 \sim 500\ km/s/Mpc$ \cite{MacCallum:2015aaa}, which was because of incorrect cosmological distance measurements that he used in the analysis. As the observations improved the value of the Hubble constant was constrained between values $40 - 100\ km/s/Mpc$ in the 1980's and 90's \cite{Freedman:2021ahq}. Today the value of the Hubble constant inferred from the late time observations using distance ladder is $H_0=73.30\pm 1.04\ km/s/Mpc$ \cite{Riess:2021jrx} and calculated from the early Universe cosmic microwave background (CMB) data is $H_0=67.37\pm 0.54\ km/s/Mpc$ \cite{Planck:2018vyg}, which are separated with more than $5\sigma$ \cite{Freedman:2021ahq}. This is called the Hubble crisis \cite{Verde:2019ivm} and there have been numerous proposals to resolve this crisis \cite{Verde:2019ivm,Vagnozzi:2019ezj,Knox:2019rjx,Riess:2019qba,DiValentino:2021izs,Perivolaropoulos:2021jda,Schoneberg:2021qvd,Shah:2021onj,Abdalla:2022yfr}. 
If the problem lies in the late time result, it could be due to the excess systematic errors in the SH0ES result \cite{Riess:2021jrx}, such as the excess brightness, known as the ``crowding problem,'' in the observations of Cepheid variables. However, this possibility is ruled out by the recent JWST's observations \cite{Riess:2024ohe,Anand:2024nim,Li:2024yoe,Freedman:2023jcz}. Since the value of the Hubble constant obtained from the late time observations seem to be upheld with further observations by JWST, 
the problem could be in the model dependent calculation of the Hubble constant from the Planck data \cite{Planck:2018vyg}. This could be due to a systematic effect in the analysis of CMB data which might come from excess lensing by surprisingly large galaxies at high redshift \cite{McGaugh:2023nkc}. Perhaps not only new early-time physics, but together also new late-time physics is required \cite{Vagnozzi:2023nrq} to resolve the Hubble tension. In this work, we study ramifications of a new model with a modified Friedmann equation \cite{Deliduman:2023caa} that changes the time-redshift relation throughout the Universe's history to resolve various cosmological problems.

Conflict of observational cosmology with the standard $\La$CDM model of the Universe is not confined only to the value of the Hubble constant. James Webb Space Telescope's (JWST) observations since its launch have shown us that there could be very massive and very large galaxies \cite{McGaugh:2023nkc,Naidu:2022wia,Castellano:2022wia,Labbe:2022ahb,Menci:2022wia,Robertson:2022gdk,Curtis-Lake:2023,ArrabalHaro:2023,Yung:2023bng,Forconi:2023izg}, and massive quasars \cite{Maiolino:2023zdu,Bogdan:2023ilu,Maiolino:2023bpi,Natarajan:2023rxq,Furtak:2023ege,Pacucci:2023oci,Ilie:2023aqu} very early in the history of the Universe, conflicting expectations of the $\La$CDM model \cite{Yung:2023bng,Forconi:2023izg,McGaugh:2023nkc,Maiolino:2023zdu,Ilie:2023aqu,Boylan-Kolchin:2022kae,Lovell:2022bhx,Gimenez-Arteaga:2022ubw,Cameron:2023zhm,Melia:2023dsy,Parashari:2023cui,Hirano:2023auh,Adil:2023ara,Menci:2024rbq,Sun:2023wqq,vanPutten:2023ths,Forconi:2023hsj,Laursen:2023tbg,Davari:2023tam}. This so-called ``impossibly early galaxy problem'' \cite{Melia:2023dsy,Melia:2014cva} require too rapid star formation in the earliest galaxies than appears to be permitted by the $\La$CDM model. For the galaxies to form \cite{Salvaterra:2010nb,Jaacks:2012rn} from the initial halo formation to the earliest population III stars to the observed sizes and masses, there is just not enough time in the $\La$CDM model \cite{Forconi:2023izg,Boylan-Kolchin:2022kae,Lovell:2022bhx,Gimenez-Arteaga:2022ubw,Melia:2023dsy,vanPutten:2023ths,Laursen:2023tbg,Boylan-Kolchin:2021fvy,Padmanabhan:2023esp}. 
A related problem is the mass of the earliest quasars \cite{Ilie:2023aqu,Inayoshi:2019fun}. From a ``incredibly massive seed'' that might be formed as a core collapse black hole \cite{Ilie:2023aqu} the observed oldest quasars require mass accretion on the Eddington or super-Eddington limit for a very long time. Thus, observations of earliest luminous galaxies and massive quasars is termed as the JWST mass problem. 
In fact, this might not be a high masses problem, but a ``time-compression problem''  \cite{Melia:2023dsy}: time too short for the observed large and massive structures to form from the initial seeds. A cosmological model that could allocate more time for the earliest large structures to form would be more conforming to the data than the $\La$CDM model. 

If the formation history of galaxies at high redshift cause a time-compression problem in $\La$CDM model \cite{Boylan-Kolchin:2021fvy,Bernal:2021yli,Krishnan:2021dyb}, then an alternative model would better solve this problem for the relevant era, not for any other era for which there is not such a time problem. Most early dark energy models decrease the sound horizon at recombination and thus increase the Hubble constant in order to ease or resolve the Hubble crisis \cite{DiValentino:2021izs,Forconi:2023hsj}. These models decrease the amount of time that passes between the Big Bang and the recombination. Since the excess dark component ceases to exist after the recombination such models do not change the time redshift relation of $\La$CDM after the recombination. Yet some other models that increase the age of the Universe drastically at $z=0$ \cite{Gupta:2023mgg} are at odd with the calculated ages of the oldest astronomical objects (OAO). The fact that those objects are seen at most at redshift $z\sim 13$ \cite{Robertson:2022gdk,Curtis-Lake:2023} and the absolute age of the old globular cluster M92 \cite{Ying:2023oie} disfavor a drastic deviation of the age of the Universe today from the value calculated in the $\La$CDM model.

In this work, we are going to discuss a new cosmological model that eases and perhaps resolves the time-compression problem. In a recent work \cite{Deliduman:2023caa} a new solution to the $f(R)$ gravity field equations in the anisotropic Bianchi type I background was presented. $f(R)$ theory of gravity is one of the simplest modifications \cite{Sotiriou:2008rp,DeFelice:2010aj,Nojiri:2010wj,Nojiri:2017ncd} of the Einstein's theory, which has an arbitrary function of the scalar curvature R as its Lagrangian density. Some well-known anomalies in both the early and the late Universe observational data \cite{COBE:1992syq,Kogut:1996us,Bennett:2010jb,deOliveira-Costa:2003utu,WMAP:2003ivt,WMAP:2003zzr,WMAP:2003elm,Planck:2015igc,Planck:2019kim} was the motivation to study cosmology in an anisotropic background, which resulted in a new model of cosmology that is called the $\ga\de$CDM model \cite{Deliduman:2023caa}. The observed anomalies include:
the lack of power in the CMB quadrupole moment \cite{Planck:2018vyg,Buchert:2015wwr,Schwarz:2015cma,Campanelli:2006vb,Cea:2022mtf,Campanelli:2007qn,Cea:2019gnu}, alignment of the quadrupole and octupole moments of CMB with each other and the motion of the solar system \cite{Copi:2013jna,Land:2005ad,Schwarz:2004gk}, observed power asymmetry between southern and northern hemispheres of the celestial sphere \cite{Planck:2019evm,Mukherjee:2015mma,Javanmardi:2016whx,Axelsson:2013mva,Eriksen:2007pc}, and the existence of so-called cold spots on the microwave sky \cite{Vielva:2003et,Cruz:2006sv}. These anomalies might collectively be pointing out the existence of a preferred direction in space \cite{Schwarz:2015cma,Cea:2022mtf,Luongo:2021nqh,Krishnan:2021jmh,Rodrigues:2007ny,Bridges:2007ne,Migkas:2020fza,Aluri:2022hzs}. 

An anisotropic cosmological background geometry is usually modeled by Bianchi type metrics, among which Bianchi type I metric is the simplest \cite{Campanelli:2006vb,Cea:2022mtf,Campanelli:2007qn,Cea:2019gnu,Collins:1973lda,Akarsu:2019pwn,Akarsu:2020pka,Akarsu:2021max,Tedesco:2018dbn,Amirhashchi:2018bic,Hossienkhani:2014zoa,Nojiri:2022idp}. Anisotropic expansion is possible in the case of $f(R)$ gravity \cite{Barrow:2005qv}. In contrast, in the case of general relativity with a positive cosmological constant an anisotropic Universe tends to isotropize asymptotically \cite{Starobinsky:1962,Wald:1983ky}.
We also stress that in the $\ga\de$CDM model $f(R)$ gravity field equations are not treated as effective Einstein equations \cite{Capozziello:2005mj,Nojiri:2006ri,Capozziello:2006dj,Faraoni:2018qdr}. Thus, this model relates the expansion of the Universe to the energy content of the Universe very differently compared to the standard $\La$CDM model.
In the $\ga\de$CDM model, different energy densities contribute to the Hubble parameter with different weights. Additionally, in the formula for the Hubble parameter, energy densities depend on the redshift differently than what their physical nature dictates. This new way of relating Universe's energy content to the Hubble parameter leads to a modified relation between the cosmic time and redshift. Although this model cannot include a cosmological constant component, a dark energy dynamically depending on redshift is a possibility.

We test the observational relevance of the $\gamma\delta$CDM model to the age problem by constraining its parameters with the ages of the OAO, prepared by Vagnozzi et al. \cite{Vagnozzi:2021tjv} based on galaxies and high-z quasars with redshifts up to $z\sim8$. OAO data requires a half-Gaussian for the (log-)likelihood function and thus cannot constrain the model parameters completely. Thus, we also best fit the model to cosmic chronometers (CC) Hubble data \cite{Jimenez:2001gg,Moresco:2012,Moresco:2015,Moresco:2016,Moresco:2020,Favale:2023lnp} and the Pantheon+ type Ia SNe data \cite{Brout:2022vxf,Scolnic:2021amr,Scolnic:pant} of the late Universe at low redshift. We find that, thanks to a modified time--redshift relation, the $\ga\de$CDM model has a more plausible time period at high redshift for large and massive galaxies and massive quasars to form, whereas the age of the Universe today is not modified significantly.

This paper is organized as follows: in the next section we summarize the derivation of the  $\ga\de$CDM model in the $f(R)$ gravity framework in an anisotropic background. In section \ref{method} we first present data sets with which we constrain the astrophysical and cosmological parameters and then summarize the data fitting method that we use. In section \ref{results} we present and discuss the implications of the results of the data fit analysis. Lastly in section \ref{conc} we summarize our work and scrutinize our results.


\section{Theory \label{theory}} 

\subsection{$f(R)$ gravity, field equations and background \label{fR}}

In this section we summarize the derivation of the $\ga\de$CDM model for the sake of completeness and to introduce the cosmological parameters that will be constrained in later sections through data analysis.

The $f(R)$ theory of gravity has an arbitrary function of the scalar curvature $R$ as its Lagrangian density. Therefore, its action is written as
\be \label{action}
S=-\fr1{2\kappa} \int d^4 x \sqrt{-g}f(R) + S_m
\ee
where $S_m$ denotes the action for matter fields, and the Einstein's constant in natural units (c=1) is $\kappa = 8\pi G$.

By varying the action ({\ref{action}) with respect to the metric tensor one gets the field equations as given by
\be
f_R R_{\mu\nu} -\fr12 f g_{\mu\nu} +\left( g_{\mu\nu}\Box - \nabla_\mu \nabla_\nu \right) f_R = \kappa T_{\mu\nu}
\label{field}
\ee
where the energy-momentum tensor is obtained with $T_{\mu\nu} = -\fr{2}{\sqrt{-g}} \fr{\delta \mathcal{S}_m}{\delta g_{\mu\nu}}$, and we denote $f_R = \fr{\partial f}{\partial R}$.

There can be a preferred direction in space which is hinted by several observations as summarized in the introduction.
We, therefore, use a Bianchi type I metric as the background geometry and an ellipsoidal model of the Universe as in \cite{Campanelli:2006vb,Campanelli:2007qn,Cea:2019gnu}. The Bianchi type I metric is given by
\be \label{B1}
ds^2=-dt^2+A(t)^2dx^2+B(t)^2 (dy^2+dz^2) \, ,
\ee
where $A(t)$ and $B(t)$ are the directional scale parameters. Using the relation for the physical volume in terms of $A(t)$ and $B(t)$, an average scale parameter $a(t)$ can be defined by $V(t) = A\cdot B^2 = a(t)^3$.

In terms of $A(t)$ and $B(t)$ and their time derivatives we define further parameters as
$H (t) = \fr13 ( \dot{A}/A +2\dot{B}/B )$ and 
$S (t) =\dot{A}/A - \dot{B}/B$, where $H(t)$ is the average Hubble parameter and $S(t)$ is the shear anisotropy parameter. The Hubble parameter can be expressed in terms of the average scale parameter $a(t)$ as $H(t) = \dot{a}/a$.
The shear scalar $\si^2(t)$ that quantifies the anisotropic expansion \cite{Leach2006} is related to $S^2(t)$ by the relation $S^2(t) = 3\si^2(t)$.
For the Bianchi type I spacetime (\ref{B1}) the scalar curvature is calculated to be
$R = 12H^2 +6\dot{H} +2 S^2 /3$.

In terms of $H(t)$ and $S(t)$ the field equations are given by
\beqa
\mathrm{(00)} &:&  f_R \left( 3 H^2+3 \dot{H}+\frac{2 S^2}{3} \right) -\fr12 f - 3f_{RR} H \dot{R} = -\kappa\rho \, , \label{00} \\
\mathrm{(11)} &:&  f_R \left( 3 H^2+2 H S+\dot{H}+\frac{2 \dot{S}}{3} \right) -\fr12 f 
- f_{RR} \left[ ( 2H - \fr23 S ) \dot{R} + \ddot{R} \right] - f_{RRR} \dot{R}^2 = \kappa p \, , \label{11} \\
\mathrm{(22)} &:&  f_R \left( 3 H^2-H S+\dot{H}-\frac{\dot{S}}{3} \right) -\fr12 f 
- f_{RR} \left[ ( 2H +\fr13 S ) \dot{R} + \ddot{R} \right] - f_{RRR} \dot{R}^2 = \kappa p \, , \label{22}
\eeqa 
where the matter content is approximated as a perfect fluid. In these equations we have $f_{RR} = \fr{\partial^2 f}{\partial R^2}$ and $f_{RRR} = \fr{\partial^3 f}{\partial R^3}$. 

Since there are no anisotropic pressure components, the left hand sides of $(11)$ and $(22)$ field equations should be the same.
Combining the equations (\ref{11}) and (\ref{22}) we obtain a trace free Gauss--Codazzi equation \cite{Leach2006,Banik2016} given by
\be \label{GC}
\fr{\dot{S}}S = - \left[ 3 \fr{\dot{a}}a + \fr{\dot{f}_R}{f_R} \right] \ ,
\ee
Which can be solved to determine $f_R = \fr{\partial f}{\partial R}$ and $S(t)$ as a function of the scale parameter $a(t)$ as
\be \label{anis}
f_R  =  \fr\varphi{a^{3-\varepsilon}} \quad and \quad S = \fr{s_0}{a^{\varepsilon}} \, ,
\ee
after assuming that time dependence of $S(t)$ is such that $\dot{S} = -\varepsilon HS$. In the above relations, $\varphi$ and $s_0$ are just the integration constants. For simplicity, we will set $\varphi = 1$ in the rest of this work.
The field equations then become
\beqa 
\label{iso} \kappa \rho &=& - f_R \left( 3 H^2+3 \dot{H} +\frac{2 S^2}{3} \right) + \fr12 f + 3H \dot{f}_R \, ,  \\
\label{iso1} \kappa p &=&  f_R \left( 3 H^2+\dot{H} \right) - \fr12 f - 2H \dot{f}_R - \ddot{f}_R \, .
\eeqa


\subsection{Solution to the field equations} 

Shear dissipates as $S^2 \propto a^{-6}$ in general relativity. In the $f(R) = R + \alpha R^2$ gravity, however, it is expected that shear dissipates more slowly \cite{Leach2006,Maartens:1994pb}. In a general $f(R)$ gravity, the contribution of shear to the Hubble parameter is due to $2S^2 /3$ term in (\ref{iso}). 
Thus, the parameter $\varepsilon$ of (\ref{anis}) should be less than three so that shear in $f(R)$ gravity dissipates slower than its general relativistic analogue. So we choose that
\be \label{gamma}
\varepsilon = 3 - \delta \quad \mathrm{with} \quad  0 < \delta < 1
\ee
The range of $\delta$ as chosen above is in line with the dynamical systems analysis and equation (60) of \cite{Leach2006}.

To solve the field equations (\ref{iso} \& \ref{iso1}) we first assume that the Hubble parameter has a polynomial dependence on $a$ as given by
\be
H^2 = \sum_\la \frac{h_\la}{a^\la} \, ,
\ee
where $\la \in \mathbb{R}$. 
The form of $f(a)$ can then be obtained by integrating $f_R (a)$ (\ref{anis}) and using the form of the scalar curvature R in terms of $a$. We find that
\be \label{f}
f(a) = \sum_\la \frac{\la (12-3\la)}{\la +\delta} \frac{h_\la}{a^{\la+\delta}} 
+ \fr{4(3-\delta)}{3(6-\delta)}  \fr{s_0^2}{a^{6-\delta}}- 2\La ,
\ee
where the integration constant is denoted by $2\La$. 

Perfect fluid is composed of the usual relativistic and non-relativistic components. Dust (m) is the non-relativistic component with vanishing pressure. A positive pressure radiation (r) and a negative pressure dark energy (e) compose the relativistic component. Dark energy component have equation of state parameter $\omega = \gamma/3 -1$ with $0< \gamma \le 2$, since the field equations (\ref{iso} \& \ref{iso1}) do not allow dark energy to be a cosmological constant.
Thus, the perfect fluid has total energy density and pressure as given by
\be \label{emt}
\rho = \fr{\rho_{e0}}{a^\gamma} + \fr{\rho_{m0}}{a^3} + \fr{\rho_{r0}}{a^4} \quad \mathrm{and} \quad p = (\fr\gamma3 -1) \fr{\rho_{e0}}{a^\gamma} + \fr13 \fr{\rho_{r0}}{a^4} \, ,
\ee
where $\rho_{m0}$, $\rho_{r0}$ and $\rho_{e0}$ are the present day dust, radiation and dark energy densities, respectively.

We determine possible $\la \in \mathbb{R}$ values and solve the Hubble parameter in terms of the perfect fluid densities by substituting relations (\ref{anis}, \ref{f} \& \ref{emt}) into the field equations. We find
\be \label{H2}
H^2 = H_0^2 \left[ \fr{\Omega_{e0}}{b_\gamma} \fr1{a^{\gamma - \delta}} + \fr{\Omega_{m0}}{b_3} \fr1{a^{3 - \delta}} + 
\fr{\Omega_{r0}}{b_4} \fr1{a^{4 - \delta}} + \fr{\Omega_{s0}}{1 -\delta} \fr1{a^{6 - 2\delta}} \right] 
\quad \mathrm{with} \quad \Omega_{s0} = \fr{s_0^2}{9 H_0^2} \, ,
\ee
where we used the dimensionless density parameters, defined by dividing respective densities with the critical density $\rho_c = 3H_0^2/8\pi G$. $\Omega_{s0}$ is the contribution of anisotropic shear to the Hubble parameter and $H_0$ is the Hubble constant at $z = 0$. Here the coefficients $b_n$ ($n = \gamma,3,4$) are given by
\be \label{bn}
b_n = - \left( 1+ \delta - \fr1{2n} (n -\delta)(4+\delta) \right) \, .
\ee

In the limit $\delta \rightarrow 0$ the general relativistic solution is obtained. In that case, one finds $b_\gamma = b_3 = b_4 = 1$ and that $\Omega_{e0} + \Omega_{m0} + \Omega_{r0} + \Omega_{s0} = 1$ as it should be for a flat Universe in the $\La$CDM model. In the same limit, the shear $\Omega_{s}$ diminishes with the sixth power of the average scale parameter ($1/a^{6}$) as is the case in general relativity.

The most important difference between this model and the $\La$CDM model is the existence of the $\delta$ parameter. This parameter affects both the contribution of different perfect fluid components to the Hubble parameter and also how different densities dissipate with the expansion of the Universe. 
Coefficients $b_n$ for $n = \gamma,3,4$ (\ref{bn}) depend on the value of $\delta$, and behave like weight factors that determine how much an individual perfect fluid component contributes to the Hubble parameter. Since $\delta$ exists also in the powers of redshift, contribution of the perfect fluid components to the expansion of the Universe is different from what their physical nature dictates. For example, radiation component dissipates with $a^4$, but its contribution to the expansion diminishes slower with $a^{4-\delta}$. This is also the case for the other perfect fluid components.

In some solutions of the field equations of the scalar--tensor theories similar dependence on the redshift was observed \cite{Boisseau:2010pd,Akarsu:2019pvi,Schiavone:2022wvq}. Solutions presented in these works are exact if the perfect fluid has only the dust component: radiation component cannot be included into the exact analytical solution and dark energy can only be included ``effectively.'' Other than that, in these works, contribution of dust to the expansion diminishes faster than its actual physical dynamics requires. This is contrary to our solution here in the $f(R)$ theory framework.


\section{Data and Methodology \label{method}} 

\subsection{Data}

We are going to constrain the cosmological parameters of the theoretical $\ga\de$CDM model with the late time observation data sets, which are the OAO age data set, CC Hubble data set, and the Pantheon+ SNe Ia data set. These data sets are physically distinct and uncorrelated.

\subsubsection{OAO Age data}

We use the data from the age of old astronomical objects (OAO) catalog prepared by Vagnozzi et al. \cite{Vagnozzi:2021tjv}, which is based on galaxies and high-z quasars with redshifts up to $z\sim8$. 

They obtained galaxy data primarily from the Cosmic Assembly Near-infrared Deep Extragalactic Legacy Survey (CANDELS) \cite{Grogin:2011ua}, which covers all the five observation fields \cite{Barro:2019,Santini:2015,Nayyeri:2017}, and also includes 32 early-time galaxies \cite{Simon:2005} in the range $0.12<z<1.85$. For high redshift quasars, they considered 7446 quasars in the range $3<z<5$ from the SDSS Data Release 7 quasar catalog \cite{Shen:2011}, 50 quasars identified with GNIRS near-IR spectroscopy in the redshift range $5.5 < z < 6.5$ from the GEMINI program \cite{Shen:2019}, and 15 quasars in the range $6.5 < z < 7$ from the Pan-STARRS1 survey \cite{Mazzucchelli:2017}. Additionally, the catalog includes the most distant 8 quasars with redshifts in the range $7 < z < 7.647$ \cite{Banados:2017unc,Mortlock:2011,Yang:2019,Yang:2020,Wang:2018,Wang:2021,Matsuoka:2019a,Matsuoka:2019b}.

As they construct the OAO age data, Vagnozzi et al. \cite{Vagnozzi:2021tjv} made various cuts based on the quality of selected data and the catalog is compiled by selecting the oldest objects from each redshift bin, resulting in a total of 114 oldest astronomical objects, comprising 61 galaxies and 53 quasars, along with information on their ages and redshifts.

Following \cite{Vagnozzi:2021tjv} we model the posterior (log-)likelihood function for OAO age data as half-Gaussian given by
\begin{equation}\label{eq:1}
\ln \mathcal{L}(\mathcal{I} | \theta,\mathcal{M}) = -\frac{1}{2}\sum_{i} \left\{
\begin{array}{ll}
    \Delta t_{i}^{2}(\theta)/\sigma_i^{2} & \text{if } \Delta t_{i}(\theta) < 0 \\
    0 & \text{if } \Delta t_{i}(\theta) \geq 0,
\end{array}
\right.
\end{equation}
with $ \Delta t_{i} = t(\theta,z_{i}) - (t_{obj,i} + \tau)$, where $t(\theta,z_{i})$ is the theoretical age of the Universe at redshift $z_i$ calculated from the model, $t_{obj,i}$ is the age of the OAO observed at redshift $z_i$, and $\tau$ is the so-called incubation time, which represents the minimum time elapsed after the Big Bang up to the formation of the OAO. Half-Gaussian choice of (log-)likelihood function exponentially suppresses the parameter ranges that give theoretical age of the Universe less than the age of OAO plus the incubation time. In contrast, the parameter ranges that give theoretical age of the Universe more than the age of OAO plus the incubation time are equally likely.

\subsubsection{CC Hubble data} 

We use data that are obtained through the cosmic chronometers method \cite{Jimenez:2001gg} comprising 32 data points with low redshift ($0.07 < z < 1.97$). They are given in Table 1 of \cite{Deliduman:2023caa}, complete with the corresponding references.  
For the data points taken from \cite{Moresco:2012,Moresco:2015,Moresco:2016} we follow the calculation of the covariance matrix as shown in \cite{Moresco:2020}. Since the rest of the data points in Table 1 of \cite{Deliduman:2023caa} are uncorrelated with the data taken from \cite{Moresco:2012,Moresco:2015,Moresco:2016}, we include them diagonally in the covariance matrix. These remaining data points are taken from \cite{Simon:2005,Zhang:2014,Jimenez:2003,Ratsimbazafy:2017,Stern:2010,Borghi:2022}.
The chi-squared function for the 32 H(z) measurements is then defined by
\be
\chi^2_{CC} = M^T Cov^{-1}M,
\ee
where $M=H^{obs}(z_i)-H^{th}(z_i)$ is the difference of the model prediction from the observational data, and $Cov^{-1}$ is the inverse of the covariance matrix.

\subsubsection{Pantheon+ Type Ia SNe data}

The data we use to constrain our cosmological model parameters are the Type Ia SNe distance modulus measurements from the Pantheon+ sample \cite{Brout:2022vxf,Scolnic:2021amr}. The Pantheon+ sample includes 1701 light curves of 1550 distinct Type Ia supernovae, with data samples ranging in redshift $0.001 < z < 2.26$. All these data can be found in the GitHub repositories of \cite{Scolnic:pant}. For the peculiar velocities of supernovae with low redshifts not to affect our analysis, we exclude data with redshift $z<0.01$. Therefore, we use the data in the redshift range $0.01 < z < 2.26$. The Pantheon+ sample also incorporates SH0ES \cite{Riess:2021jrx} Cepheid host distances to constrain the absolute magnitude parameter $M$. 

The distance residuals are divided into two parts \cite{Brout:2022vxf}:
\begin{equation} \label{res}
\Delta \mu_i =
\begin{cases} 
    \mu_{\text{dat},i} - \mu_i^{\text{Cepheid}} & \text{, if } i \in \text{Cepheid hosts} \\
    \mu_{\text{dat},i} - \mu_{\text{model}}(z_i) & \text{, otherwise}\ .
\end{cases}
\end{equation}
The observed distance modulus is defined by
\begin{equation}
\mu_{dat}=m_b ^{corr}-M.
\end{equation}
The corrected and standardized $m_b^{corr}$ magnitudes are provided in the Pantheon+ \texttt{`m\_b\_corr'} column \cite{Scolnic:pant}.
The model-dependent distance modulus is defined as \cite{Lovick:2023tnv}
\begin{equation}
\mu_{model}=5\log_{10}\left(\frac{d_L}{10\text{pc}}\right)
\end{equation}
where $d_L$ is the luminosity distance that includes the model parameters. It is given by \cite{Alonso-Lopez:2023hkx}
\begin{equation}
d_L=c \left(1+z_{hel}\right) \int\limits^{z_{\text{HD}}} _0\frac{dz'}{H(z')}
\end{equation}
where $z_{\text{hel}}$ is heliocentric redshift of SNe and $z_{\text{HD}}$ is the CMB frame redshift with peculiar velocity corrections. These data are taken from the  \texttt{`zhel'} and \texttt{`zHD'} columns of Pantheon+ data file \cite{Scolnic:pant}.

The absolute magnitude $M$ is a parameter that is independent of the cosmological model and depends on the dynamics of the supernova. 
There is a degeneracy between $H_0$ and $M$. We use the SH0ES Cepheid distance measurements (indicated by the `\texttt{IS\_CALIBRATOR=1}' in the data file \cite{Scolnic:pant}) to break this degeneracy. We set absolute magnitude $M$ as a parameter. The parameter prior is taken to be uniform in the range $\left[-20,-18\right]$ \cite{Lovick:2023tnv,Alonso-Lopez:2023hkx}. 
 
 We excluded data with low redshifts, $z<0.01$. We thus have 1580 Hubble Diagram data points and 10 Cepheid distance (indicated by the `\texttt{CEPH\_DIST}' in the data file \cite{Scolnic:pant}) data points remained.

The chi-squared function for the Pantheon+ data is given by:
\begin{equation}
\chi^2= \Delta \mu^TC^{-1}\Delta\mu
\end{equation}
where $\Delta \mu$ is the vector of distance modulus residuals as given in equation (\ref{res}) and $C$ is the covariance matrix which includes the statistical and systematic covariance matrices \cite{Scolnic:pant}.


\subsection{Methodology} \label{metod}

To constrain the cosmological parameters of the $\ga\de$CDM model we use the Bayesian inference method \cite{Padilla:2019, Hogg:2010}. For the CC Hubble and the Pantheon+ data sets this method finds the maximum of the likelihood function,
\be \label{Like}
\mathcal{L}(\mathcal{I} | \theta,\mathcal{M}) \propto \exp{\left[-\frac12 \chi^2 (\mathcal{I} | \theta,\mathcal{M})\right]} \ ,
\ee
where, symbolically, $\mathcal{M}$ denotes the given model, $\mathcal{I}$ the data set and $\theta$ the parameters of the model that are to be constrained. $\chi^{2} (\mathcal{I} | \theta,\mathcal{M})$ is the chi-squared function, whose minimum value maximizes the likelihood function.

In this work, we use the UltraNest\footnote{\url{https://johannesbuchner.github.io/UltraNest/}} package \cite{Buchner:2021} to derive posterior probability distributions and the maximum likelihood function with the nested sampling Monte Carlo algorithm MLFriends \cite{Buchner:2014,Buchner:2017}. Nested sampling is a Monte Carlo algorithm that computes an integral over the model parameters.

We perform the Bayesian analysis of the $\ga\de$CDM model first with the age data of OAO \cite{Vagnozzi:2021tjv}, then OAO age data together with CC Hubble data, and finally OAO age data together with  CC Hubble and Pantheon+ data sets. This way we analyzed whether the parameters of the model remain in the same range for the distinct uncorrelated late time data sets, and whether the fitted value of the Hubble constant $H_0$ has a closer value to SH0ES result \cite{Riess:2021jrx} or the Planck result \cite{Planck:2018vyg}. For the joint data sets the likelihood function is a multivariate joint Gaussian likelihood given by (\ref{Like}). We also perform the Bayesian analysis of the $\La$CDM model with the same data sets to decide whether our model has more relevant results compared to the standard model of cosmology.

$\ga\de$CDM model has five cosmological parameters given by $H_0,\ \Omega_{m0},\ \Omega_{s0},\ \gamma$ and $\delta$ (\ref{H2}), and OAO age data needs two extra astrophysical parameters. The parameter $\tau$ is the so called incubation time, which determines the time that passes from Big Bang until the formation of oldest elliptical galaxies whose ages are included in the data set. Whereas the parameter $z_f$ is the redshift at which the oldest quasars are typically formed. There are expectations from astrophysical observations on the range of values for both $\tau$ and $z_f$, however, we opted to determine them through the Bayesian analysis. The constrained values of these parameters will be another test of the theoretical models. In a cosmological model, if the value of incubation time, $\tau$, turns out to be too small, or the redshift $z_f$ turns out to be too large so as to be in conflict with the astrophysical expectations about the time scale of the structure formation, then that model should be less relevant to describe the history of the Universe. 

For the radiation density today, $\Omega_{r0}$, we use the relation $\Omega_{r0}=h^{-2} (2.469\times 10^{-5})(1+\frac{7}{8}(\frac{4}{11})^{4/3}N_{eff})$ with $N_{eff}=3.046$ as given in \cite{Mukhanov:2005sc} determined by the Stefan-Boltzmann law and the value of CMB temperature. We fix the present value of the dark energy density, $\Omega_{e0}$, in terms of other parameters at $a(t_0) = a_0 = 1$ and then constrain five cosmological and two astrophysical parameters via the Bayesian analysis. 

We choose the priors in order to scan the parameter space throughly. For the parameters $h=H_0/100$ and $\Omega_{m0}$, uniform prior distributions are chosen, given by $0.55 \leqslant h \leqslant 0.85$ and $0.2 \leqslant \Omega_{m0} \leqslant 0.4$, respectively. For $\Omega_{s0}$, we choose logarithmic prior distribution given by $-16 \leqslant \log_{10} \Omega_{s0} \leqslant 0.0$ due to its expected very low value.
For the analysis summarized in Table \ref{tmv} logarithmic prior distribution is chosen for $\gamma$ with $0.001 \leqslant \gamma \leqslant 3.0$ and uniform prior distribution is chosen for $\delta$ with $0.0 \leqslant \delta \leqslant 0.772$. The chosen values for the prior bounds on model parameters are explained in \cite{Deliduman:2023caa}. 

For the oldest elliptical galaxies the prior for the incubation time $\tau$ is chosen to be Gaussian centered at $0.2\ Gyr$ with standard deviation $0.1\ Gyr$. Vagnozzi et al. \cite{Vagnozzi:2021tjv} use the incubation time distribution derived in \cite{Jimenez:2019onw} and presented in \cite{Valcin:2020vav} as prior (so-called J19 prior) for the incubation time parameter $\tau$. As pointed in \cite{Wei:2022plg}, J19 prior depends ``weakly'' on the $\La$CDM model and thus we opted for a simple Gaussian prior for the parameter $\tau$.
Use of the Gaussian prior or the J19 prior does not affect the results in an important way \cite{Costa:2023cmu}.
Since our aim in this paper is to analyze whether our $\gamma\delta$CDM model has anything new and different to say compared to the $\Lambda$CDM model, we do not use radically different prior for the incubation time. Otherwise, the comparison of our results with the ones reported in refs. \cite{Vagnozzi:2021tjv, Wei:2022plg,Costa:2023cmu} would be very difficult.
It is possible that for the low redshift galaxies in the OAO data the incubation times are larger.
This possibility is discussed previously in \cite{Vagnozzi:2021tjv}. If such is the case, this will change the posterior bounds on $H_0$: ``more stringent, yet less conservative, limits'' \cite{Vagnozzi:2021tjv}. Nevertheless, we can justify the prior chosen for the incubation time in the following way: J19 prior is obtained from observation of globular clusters with very-low-metallicity stars \cite{Jimenez:2019onw}. These are the oldest stars in a galaxy and their incubation time gives a plausible estimate of the incubation time for the galaxy. Additionally, theoretical considerations indicate that the local dynamics determines the  timescale for star formation in galaxies and that is on the order of $100\ Myr$ \cite{Laursen:2023tbg}, which agrees with the prior we choose.

For the oldest quasars data set, we calculate the incubation time as the age of the Universe at the quasar formation redshift $z_f$. For the quasar formation redshift, uniform prior is chosen in the range $10 \leqslant z_f \leqslant 30$ \cite{Jimenez:2019onw,CurtisLake:2023}. Thus in total seven parameters are constrained with OAO age and CC Hubble data sets: two astrophysical parameters ($\tau$ and $z_f$) and five cosmological parameters ($H_0,\ \Omega_{m0},\ \Omega_{s0},\ \gamma$ and $\delta$). For the data sets that include Pantheon+ SNe Ia data we also constrain the supernova absolute magnitude parameter $M$.
The prior for this parameter is taken to be uniform in the range $-20\leqslant M\leqslant-18$ \cite{Lovick:2023tnv,Alonso-Lopez:2023hkx}.


\section{Results and Discussion \label{results}} 

The age of the Universe at any redshift can be calculated by evaluating the integral,
\be \label{agez}
t_U (z) = \int\limits_z^\infty \frac{dz'}{(1+z')H(z')}\ ,
\ee
that includes the Hubble parameter (\ref{H2}).
Any object, observed at a particular redshift $z$, is inside the Universe and thus its age calculated via independent means should be less than the age of the Universe at redshift $z$. This provides a constraint \cite{Vagnozzi:2023nrq} on the possible models of the cosmological expansion, since the age of the Universe depends on the form of the Hubble parameter $H(z)$ (\ref{agez}). As explained in Section \ref{metod}, this constraint is formulated in the Bayesian analysis as half-Gaussian posterior (log-)likelihood function (\ref{eq:1}) for the age data of OAO.

\begin{figure*}[hbt!]
\centering
\begin{subfigure}
    \centering 
    \includegraphics[width=0.49\textwidth]{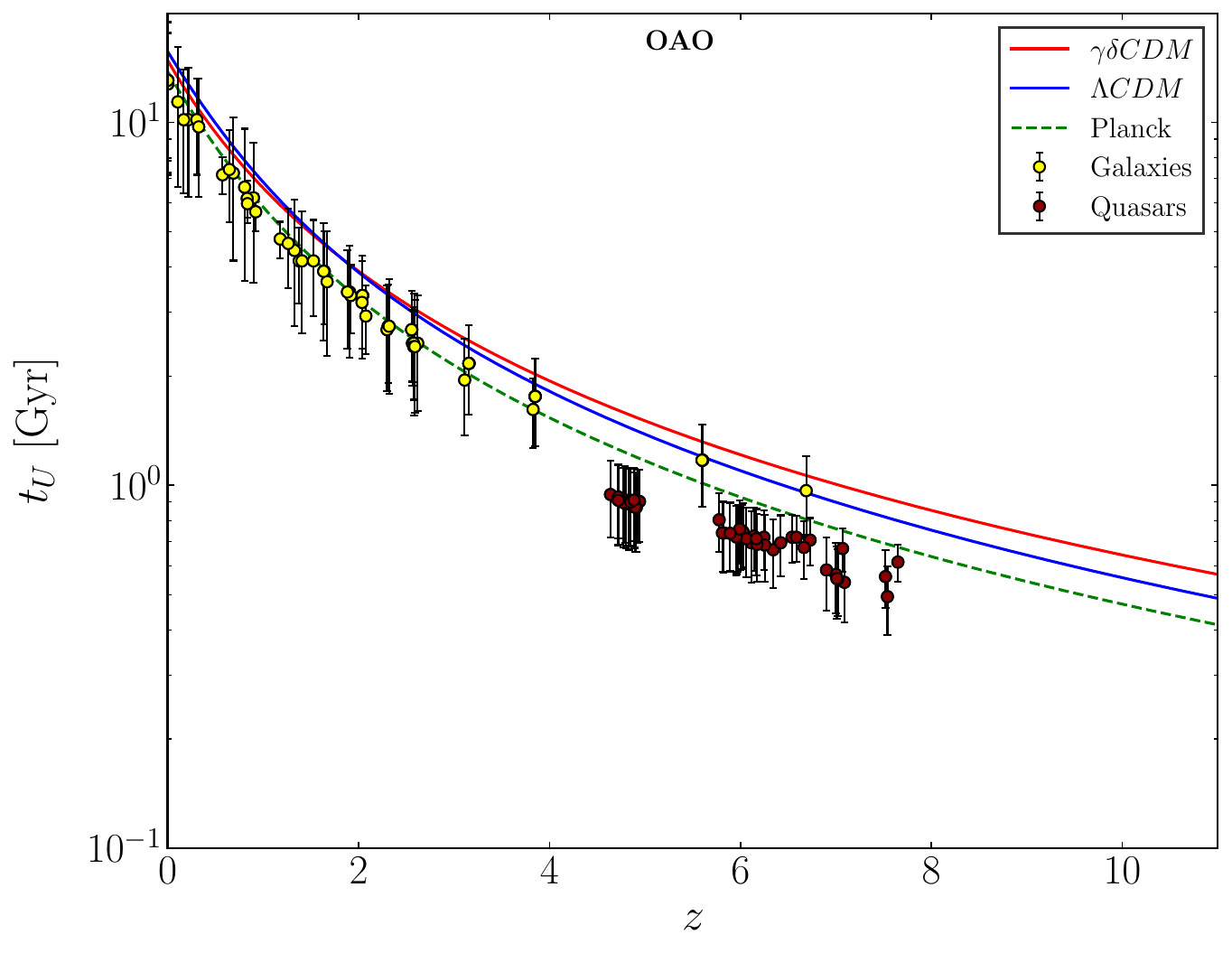}
    \label{fig:sub1a}
\end{subfigure}
\begin{subfigure}
    \centering 
    \includegraphics[width=0.49\textwidth]{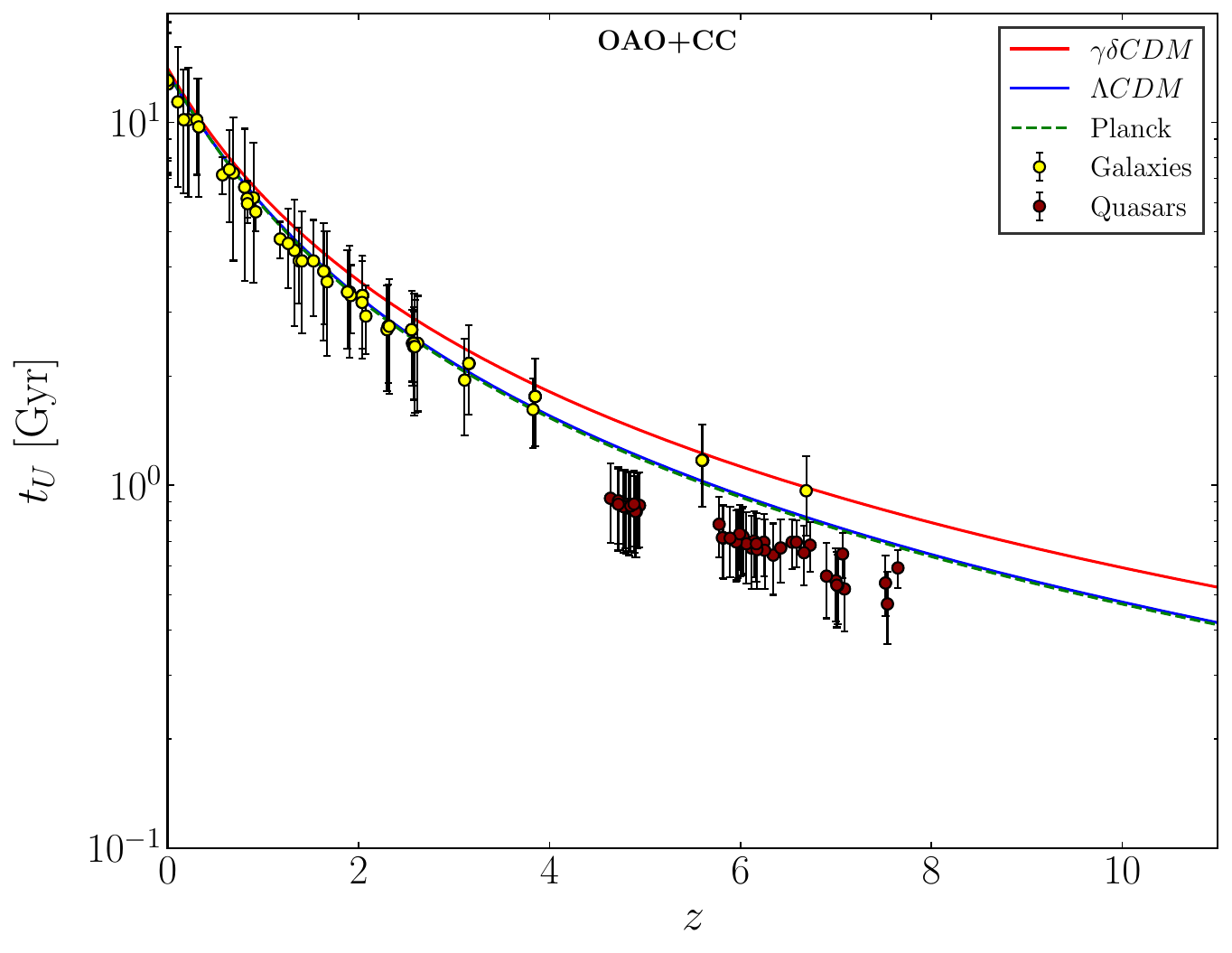}
    \label{fig:sub1b}
\end{subfigure}
\begin{subfigure}
    \centering 
    \includegraphics[width=0.49\textwidth]{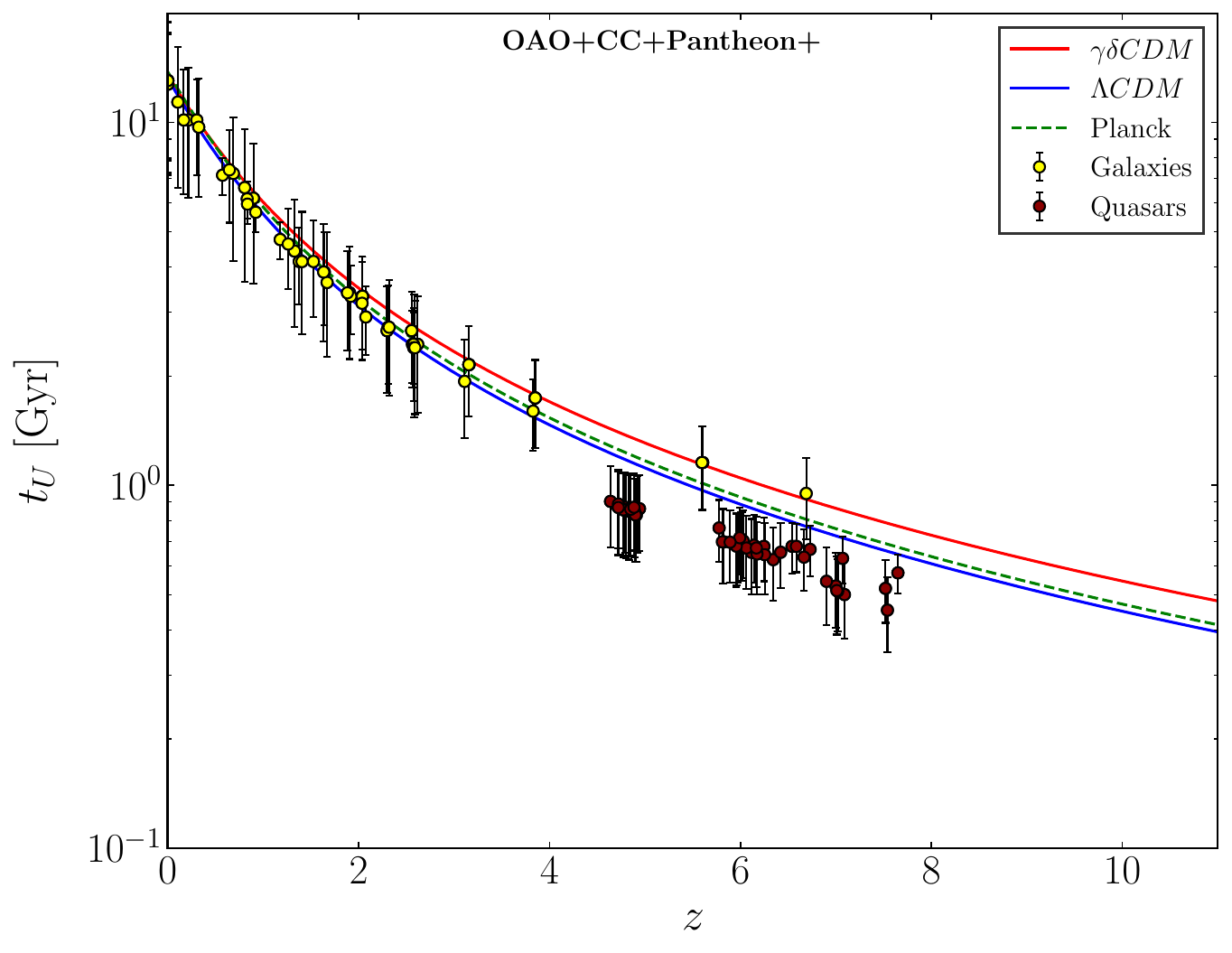}
    \label{fig:sub1d}
\end{subfigure}
\begin{subfigure}
    \centering 
    \includegraphics[width=0.49\textwidth]{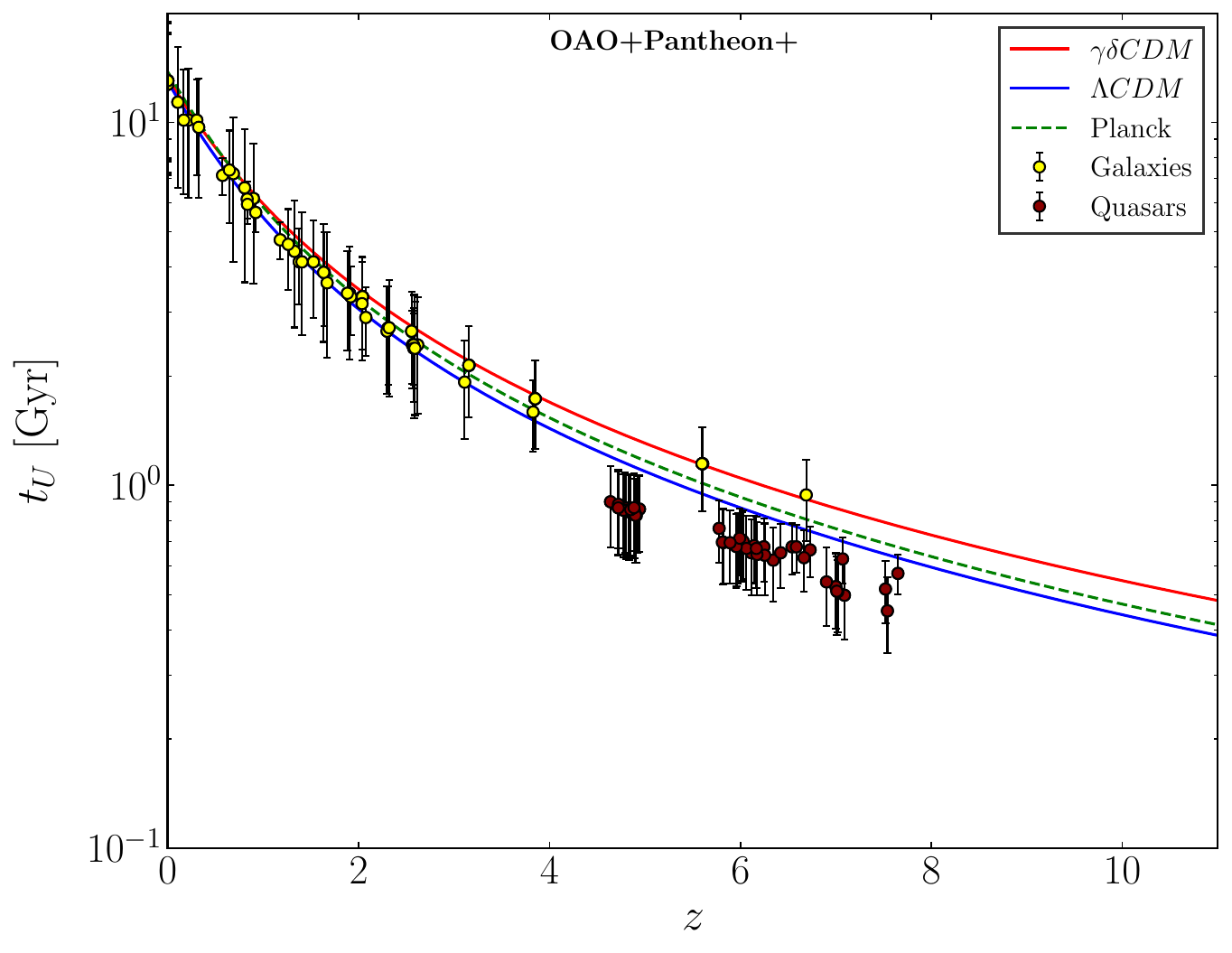}
    \label{fig:sub1c}
\end{subfigure}
\caption{The age-redshift diagram for the OAO, and the $\gamma\delta$CDM and $\Lambda$CDM cosmological models. Red and blue curves are drawn with the posterior values of the respective model parameters, obtained after a Bayesian analysis. Green dashed curve is the age-redshift relation in the $\Lambda$CDM model with the Planck collaboration values \cite{Planck:2018vyg}. Clockwise from top-left we have age-redshift diagrams for the OAO, OAO+CC, OAO+Pan+ and OAO+CC+Pan+ data sets.}
\label{fzf}
\end{figure*}

We performed the Bayesian analysis of the $\ga\de$CDM model with the age data of OAO \cite{Vagnozzi:2021tjv}, as well as OAO age data together with CC Hubble data and also together with Pantheon+ SNe Ia data, and finally OAO age data together with CC Hubble and Pantheon+ data sets. This analysis allowed us to see whether the parameters of the model remain in the same range for the distinct uncorrelated late time data sets, and whether the fitted value of the Hubble constant $H_0$ has a closer value to SH0ES result \cite{Riess:2021jrx} or the Planck result \cite{Planck:2018vyg}. For the joint data sets, the likelihood function is a multivariate joint Gaussian likelihood given by (\ref{Like}). We also performed the Bayesian analysis of the $\La$CDM model with the same data sets to decide whether our model has more relevant results compared to the standard model of cosmology.

We present the age-redshift diagrams for the OAO, together with $\gamma\delta$CDM and $\Lambda$CDM cosmological models in Fig. \ref{fzf}. In these diagrams red and blue curves are drawn with the posterior values of the respective model parameters, obtained after the Bayesian analysis. Furthermore, age-redshift relation in the $\Lambda$CDM model with the Planck collaboration values \cite{Planck:2018vyg} is drawn with the green dashed curve. We have, clockwise from top-left, the age-redshift diagrams for the OAO, OAO+CC, OAO+Pan+ and OAO+CC+Pan+ data sets. 
We remark that in all the data sets the $\gamma\delta$CDM model fits better the data sets than the $\Lambda$CDM model. We note that the most prominent difference between the age-redshift relations of two cosmological models is the OAO+CC data set. These two data sets complement each other well: OAO age data set includes absolute ages of the oldest astronomical objects, whereas CC Hubble data depend on the differential age measurements \cite{Vagnozzi:2023nrq}. Absolute age and differential age measurements are uncorrelated and the $\gamma\delta$CDM model fits better to this combined data set compared to the $\Lambda$CDM model. 
This observation is corroborated by the posterior values of the model parameters obtained with the Bayesian analysis.

\begin{table*}[hbt!]
\centering
\def\arraystretch{1.5}
\begin{tabular}{|l|c|c|c|c|}
\hline 
\hline 
\textbf{Data set} & OAO & OAO+CC & OAO+Pan+ & OAO+CC+Pan+ \\
\hline 
$\gamma\delta$CDM   \\
\hline 
$\mathbf{H_0}$ & $62.06^{+8.03}_{-4.98}$ & $66.02^{+4.19}_{-4.04}$
 & $72.65^{+2.78}_{-2.73}$ & $71.00^{+1.88}_{-1.92}$  \\
$\mathbf{\Omega_{m0}}$ & $0.266^{+0.073}_{-0.049}$  & $0.267^{+0.058}_{-0.044}$
 & $0.246^{+0.037}_{-0.030}$ & $0.254^{+0.034}_{-0.033}$  \\ 
$\log_{10}(\mathbf{\Omega_{s0}})$ & $< -10.18$ & $< -10.42$ & $< -10.51$ & $< -10.38$ \\ 
$\mathbf{\gamma}$ & $1.023^{+0.936}_{-0.697}$  & $0.846^{+0.586}_{-0.502}$
 & $0.550^{+0.195}_{-0.230}$ & $0.507^{+0.212}_{-0.229}$ \\ 
$\mathbf{\delta}$ & $0.174^{+0.225}_{-0.132}$  & $0.151^{+0.158}_{-0.107}$
 & $0.115^{+0.084}_{-0.075}$ & $0.104^{+0.089}_{-0.070}$ \\ 
$\mathbf{\tau}\ (Gyr)$ & $0.173^{+0.096}_{-0.092}$  & $0.172^{+0.091}_{-0.093}$
 & $0.147^{+0.088}_{-0.091}$ & $0.156^{+0.088}_{-0.091}$ \\ 
$\mathbf{z_f}$ & $20.63^{+6.47}_{-6.87}$ & $20.72^{+6.40}_{-6.63}$
 & $20.72^{+6.37}_{-6.61}$ & $20.47^{+6.48}_{-6.35}$ \\ 
$\mathbf{M}$ &  &  & $-19.266^{+0.082}_{-0.084}$ & $-19.317^{+0.056}_{-0.059}$ \\ 
$\mathbf{min\ \chi}^2$ & $0.0$ & $14.50$ & $1388$ & $1405$ \\ 
\hline 
$\Lambda$CDM   \\
\hline 
$\mathbf{H_0}$ & $62.45^{+8.34}_{-5.37}$ & $69.74^{+3.81}_{-3.73}$
 & $71.18^{+2.58}_{-2.55}$ & $69.87^{+1.83}_{-1.83}$ \\
$\mathbf{\Omega_{m0}}$ & $0.261^{+0.070}_{-0.044}$ & $0.286^{+0.052}_{-0.046}$
 & $0.323^{+0.017}_{-0.017}$ & $0.320^{+0.017}_{-0.016}$  \\ 
$\mathbf{\tau}\ (Gyr)$ & $0.164^{+0.093}_{-0.093}$ & $0.120^{+0.077}_{-0.083}$
 & $0.062^{+0.077}_{-0.062}$ & $0.078^{+0.074}_{-0.073}$ \\ 
$\mathbf{z_f}$ & $20.70^{+6.23}_{-6.63}$ & $20.82^{+6.25}_{-6.11}$
 & $21.25^{+5.98}_{-5.89}$ & $21.09^{+6.17}_{-5.99}$ \\ 
$\mathbf{M}$ &  &  & $-19.319^{+0.076}_{-0.079}$ & $-19.360^{+0.054}_{-0.056}$ \\ 
$\mathbf{min\ \chi}^2$ & $0.0$ & $14.50$ & $1390$ & $1407$ \\ 
\hline 
\hline
\end{tabular} 
\caption{The posterior constraints with 68\% confidence level on the free cosmological ($H_0,\ \Omega_{m0},\ \Omega_{s0},\ \gamma$ and $\delta$) parameters of the $\ga\de$CDM model and also astrophysical ($\tau, z_f$ and $M$) parameters for different data set combinations, together with the posterior constraints on the parameters of the $\La$CDM standard model. 
Minimum values of the chi-squared functions for both models are also presented.
$H_0$ has units of $km/s/Mpc$.} 
\label{tmv}
\end{table*}

We present the posterior constraints with 68\% confidence level on the free cosmological ($H_0,\ \Omega_{m0},\ \Omega_{s0},\ \gamma$ and $\delta$) parameters of the $\ga\de$CDM model, and also astrophysical ($\tau, z_f$ and $M$) parameters for different data set combinations in Table \ref{tmv}. We also present the corresponding likelihood and contour plots for 2D posterior distributions, with 1$\sigma$ and 2$\sigma$ confidence regions, for the free parameters in Fig. \ref{ftr}, together with 1D posterior distributions. 
Additionally, we combine the likelihood and contour plots for 2D posterior distributions for the different data sets in Fig. \ref{ftrc} to simplify the comparison of the results.
These plots are produced by GetDist \cite{Lewis:2019xzd}. Moreover, we provide marginalized constraints for the parameters of the $\La$CDM standard model (for which there are only $H_0,\ \Omega_{m0}$, together with $\tau, z_f, M$) in Table \ref{tmv} again for comparison purposes. 
Lastly, Table \ref{tmv} also includes the minimum values of the combined $\chi^2$ functions, obtained via Bayesian analysis, for the different data sets for both of the models. In the case of analysis with only the OAO age data, the minimum of $\chi^2$ function turns out to be vanishing for both models due to the half-Gaussian form of the likelihood. In the cases of data set combinations, $\ga\de$CDM model has lower minimum $\chi^2$ values compared to the reference $\La$CDM model.

In the case of Bayesian analysis with only OAO age data and OAO plus CC data sets the 1$\sigma$ regions of Hubble constant are very large and these data sets do not constrain the models effectively. This has been also commented on in \cite{Vagnozzi:2023nrq,Costa:2023cmu}. The reason of the poor fit is that in the Bayesian analysis the posterior (log-)likelihood function is half-Gaussian (\ref{eq:1}) for the OAO age data and the error bars for some CC Hubble data points are very large. Nevertheless, the upper 1$\sigma$ bound on $H_0$ is closer to the low-redshift value \cite{Riess:2021jrx} than the high-redshift value \cite{Planck:2018vyg}. 
We note that large uncertainties in the OAO data set is expected to cause the inferred posterior values of the cosmological parameters to be less consistent with the actual value of those cosmological parameters, as demonstrated with a mock data in \cite{Costa:2023cmu}. As the uncertainties of the age measurements are made smaller with further observations, the cosmological models will be better constrained and the posterior bounds on cosmological parameters will get smaller as well.

When we include more data, the quality of the fit increases drastically. Both OAO plus Pantheon+ data set and all three data sets combined constrain the model parameters with smaller 1$\sigma$ bounds. As seen in Fig. \ref{Mv}, for both data sets, the 
1$\sigma$ region of the posterior value of the Hubble constant in the $\ga\de$CDM model and the 
1$\sigma$ region of the local value of the Hubble constant \cite{Riess:2021jrx} overlap. The effective matter density in $\gamma\delta$CDM model is calculated from Table \ref{tmv} to be 
$\Omega_{m0}/b_3 = 0.285^{+0.043}_{-0.035}$ in the case of OAO+Pan+ data set, and $\Omega_{m0}/b_3 = 0.290^{+0.038}_{-0.038}$ in the case of OAO+CC+Pan+ data set. Thus, the OAO age data together with the SNe Ia brightness data constrain the parameters of the $\gamma\delta$CDM model consistent with the value of the Hubble constant obtained by SH0ES collaboration via distance ladder \cite{Riess:2021jrx}.

\begin{figure*}[hbt!]
\centering
\begin{subfigure}
    \centering 
    \includegraphics[width=0.49\textwidth]{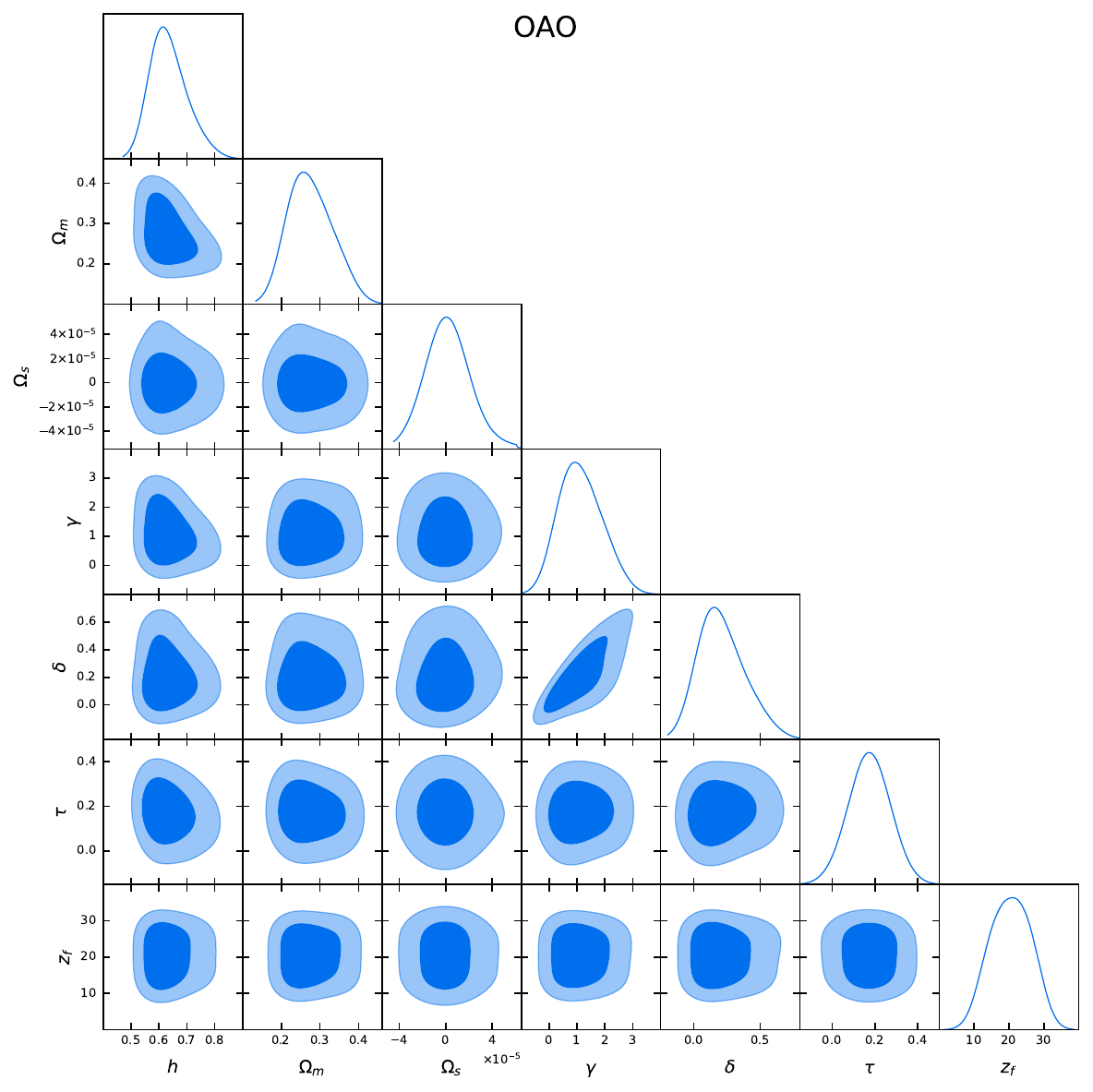}
    \label{fig:sub2a}
\end{subfigure}
\begin{subfigure}
    \centering 
    \includegraphics[width=0.49\textwidth]{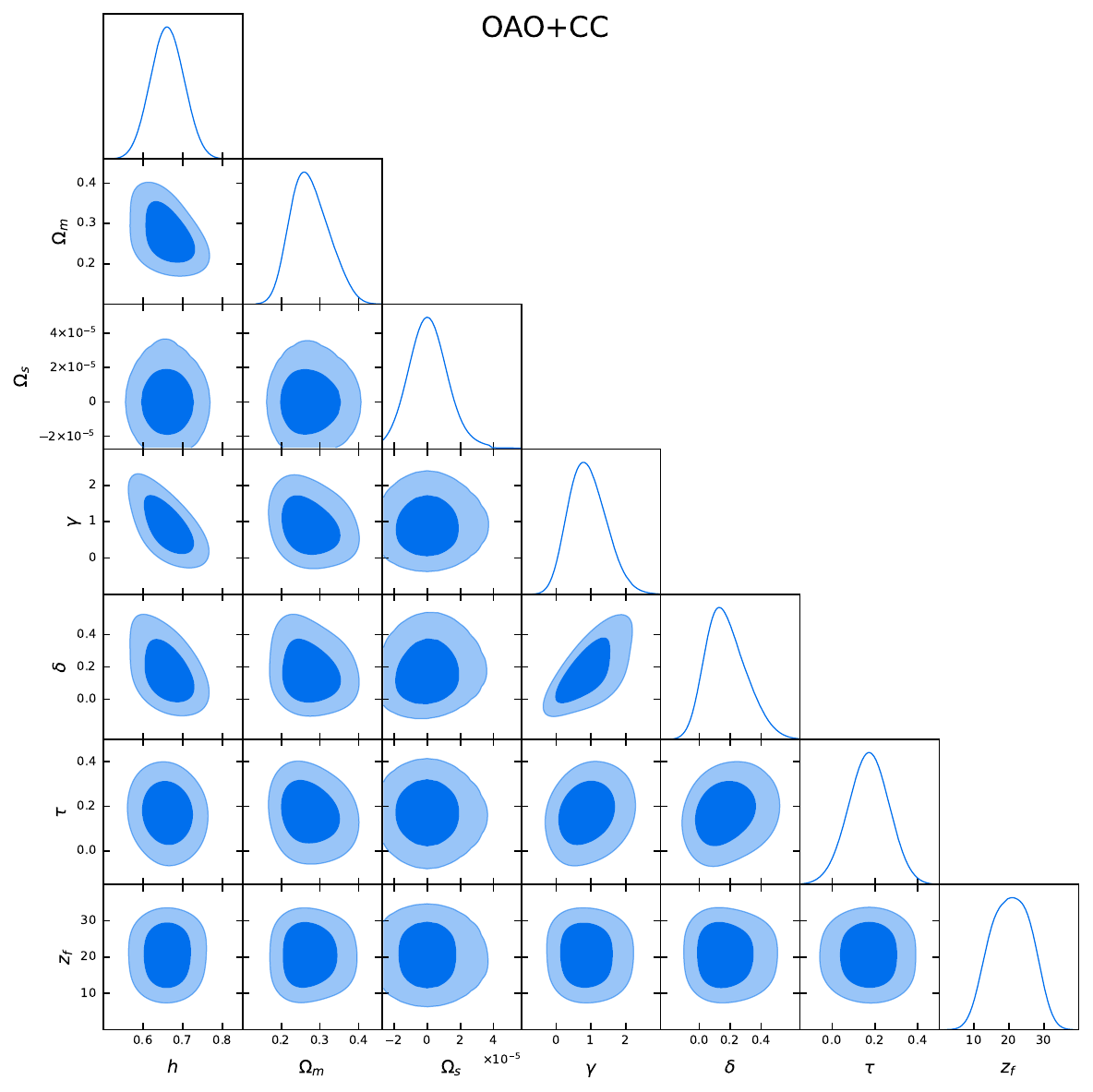}
    \label{fig:sub2b}
\end{subfigure}
\begin{subfigure}
    \centering 
    \includegraphics[width=0.49\textwidth]{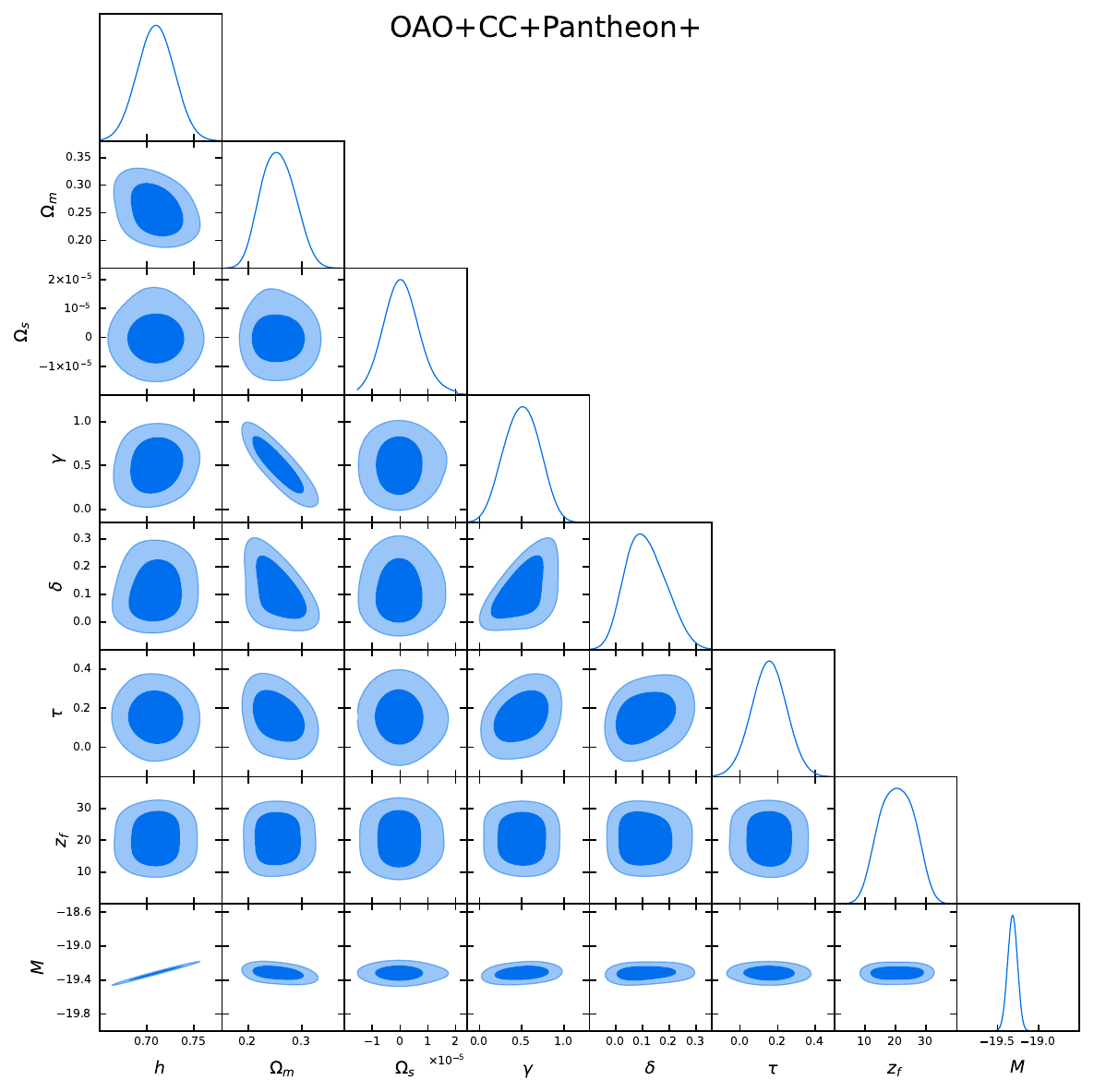}
    \label{fig:sub2d}
\end{subfigure}
\begin{subfigure}
    \centering 
    \includegraphics[width=0.49\textwidth]{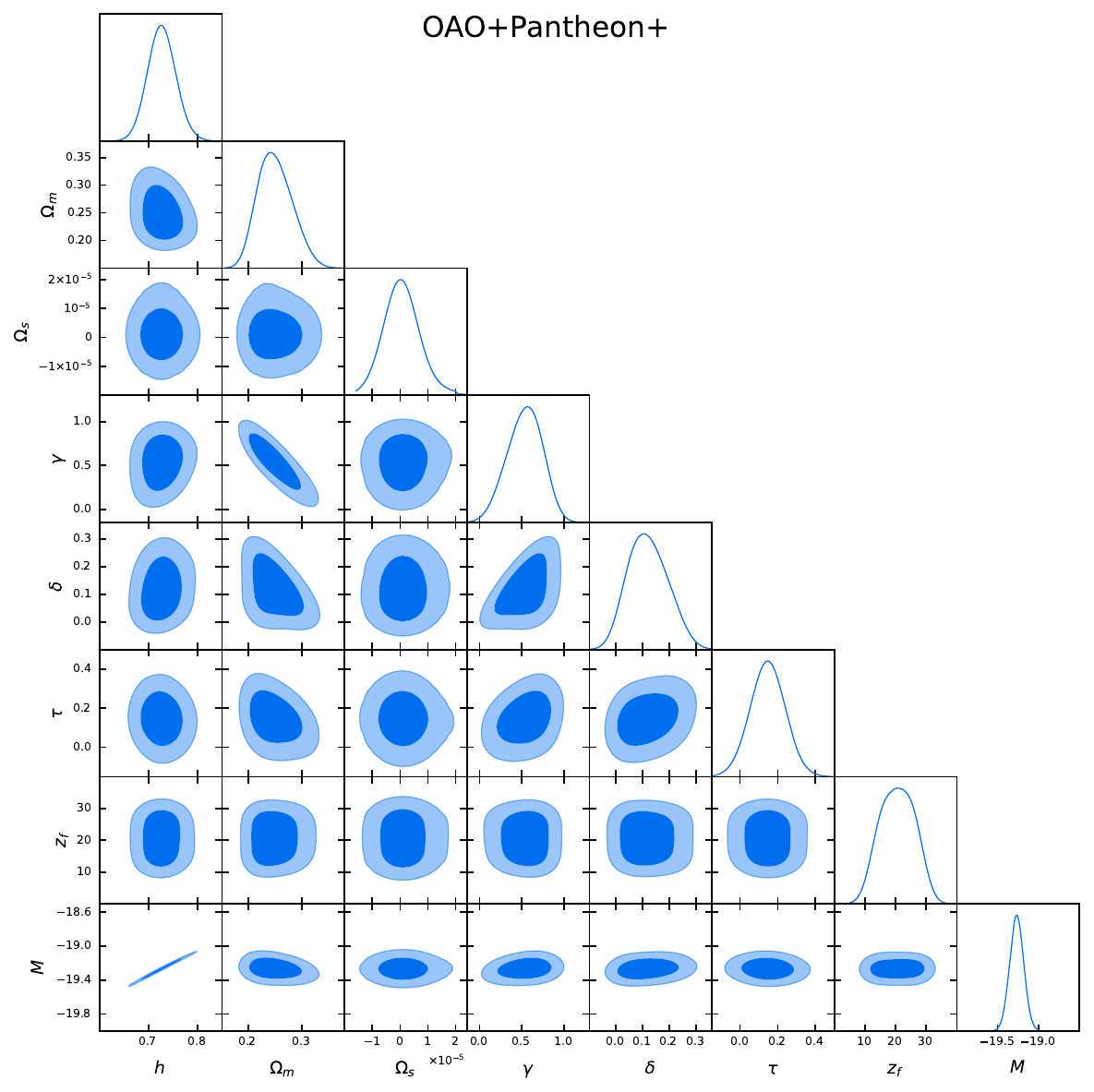}
    \label{fig:sub2c}
\end{subfigure}
\caption{The likelihood and contour plots for 2D joint posterior distributions, with 1$\sigma$ and 2$\sigma$ confidence regions, for the free cosmological ($H_0,\ \Omega_{m0},\ \Omega_{s0},\ \gamma$ and $\delta$) parameters of the $\ga\de$CDM model and also astrophysical ($\tau, z_f$ and $M$) parameters for different data set combinations, together with 1D marginalized posterior distributions. Clockwise from top-left we have 1D and 2D marginalized posterior distributions for the OAO, OAO+CC, OAO+Pan+ and OAO+CC+Pan+ data sets.}
\label{ftr}
\end{figure*}

\begin{figure*}[hbt!]
\centering
\includegraphics[width=0.95\textwidth]{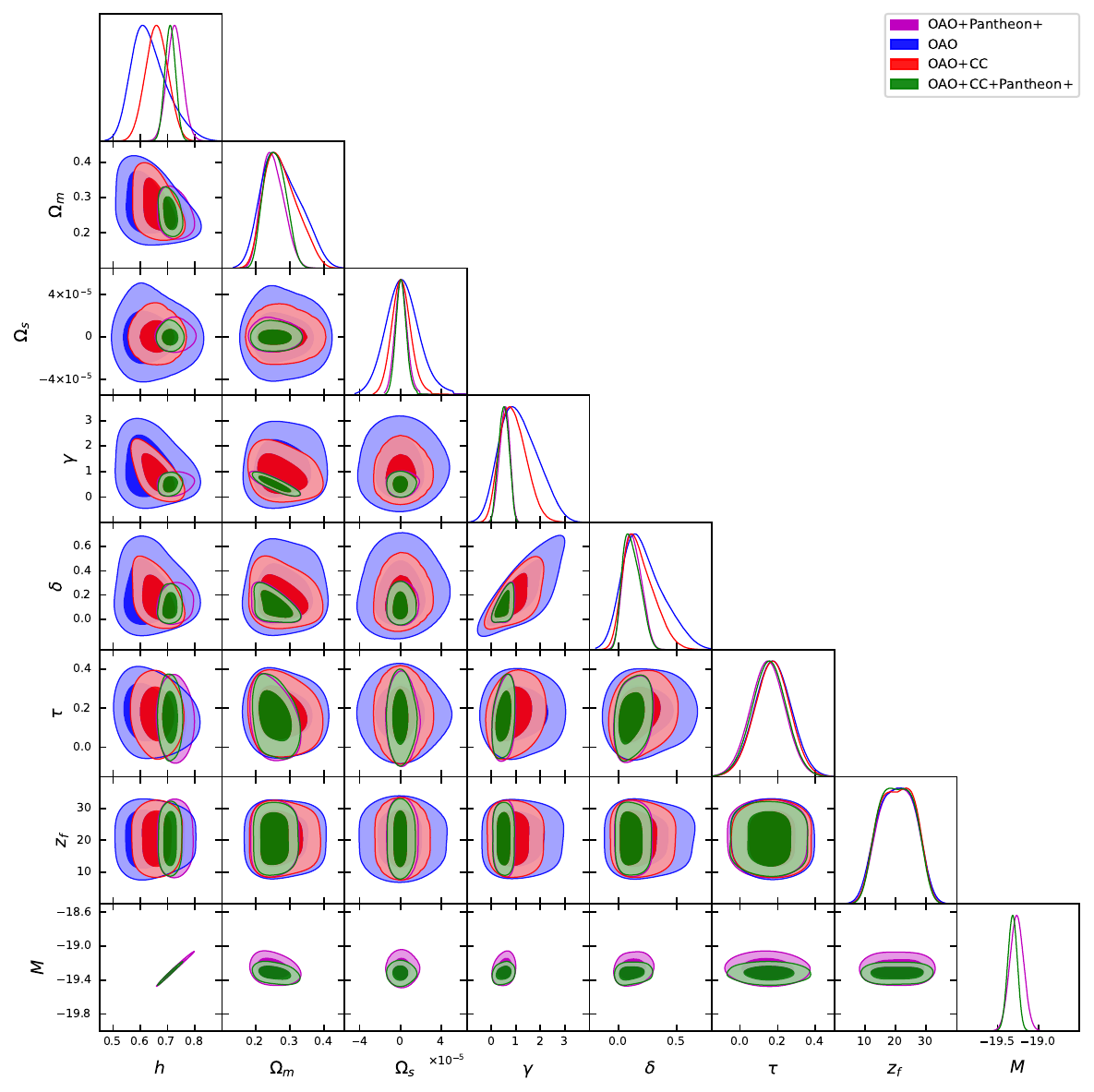}
\caption{Combined likelihood and contour plots for 2D joint posterior distributions, with 1$\sigma$ and 2$\sigma$ confidence regions, for the free cosmological ($H_0,\ \Omega_{m0},\ \Omega_{s0},\ \gamma$ and $\delta$) parameters of the $\ga\de$CDM model and also astrophysical ($\tau, z_f$ and $M$) parameters for different data set combinations, together with 1D marginalized posterior distributions.}
\label{ftrc}
\end{figure*}

\begin{figure*}[hbt!]
\centering
\begin{subfigure}
    \centering 
    \includegraphics[width=0.49\textwidth]{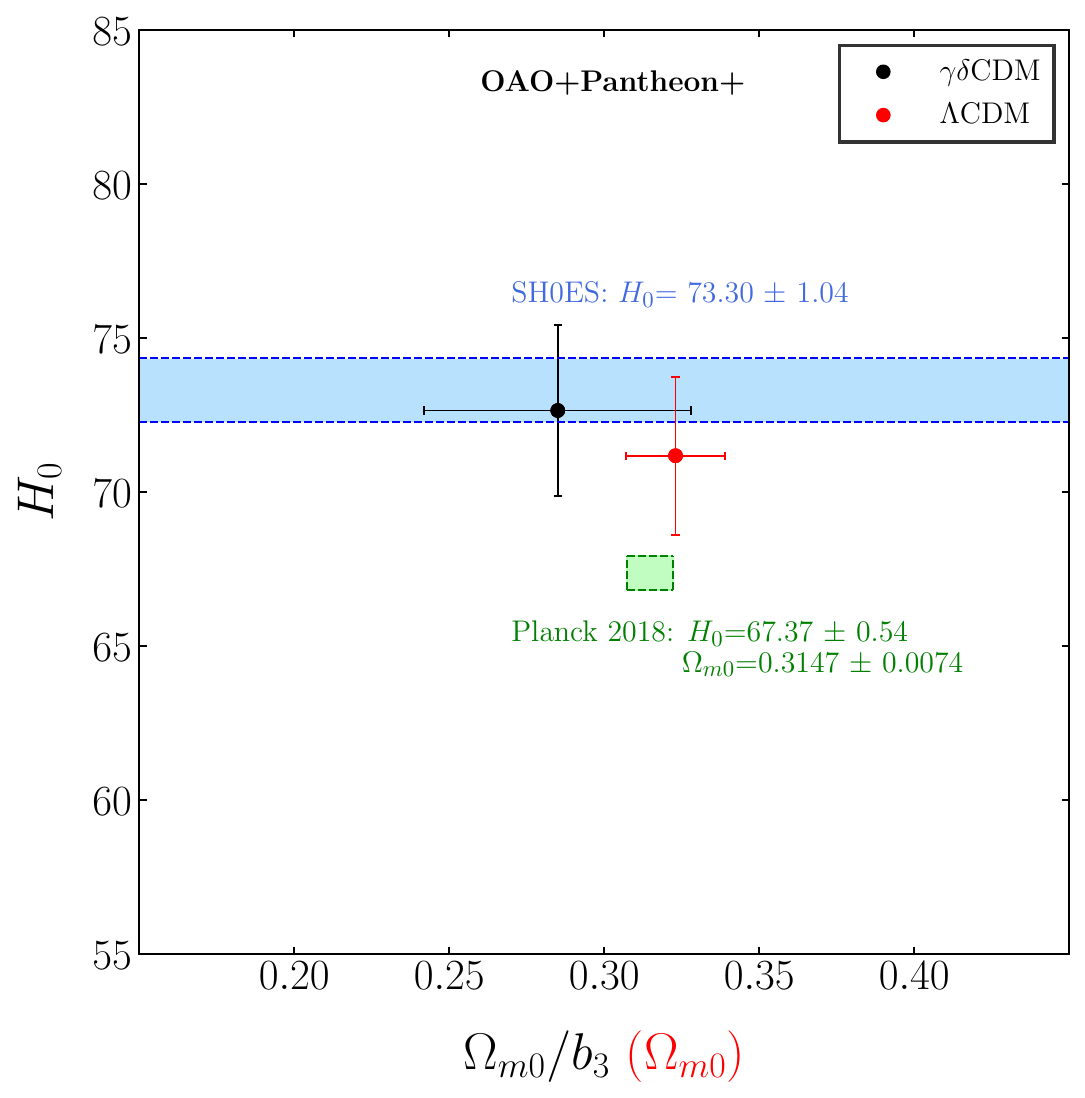}
    \label{fig:sub3a}
\end{subfigure}
\begin{subfigure}
    \centering 
    \includegraphics[width=0.49\textwidth]{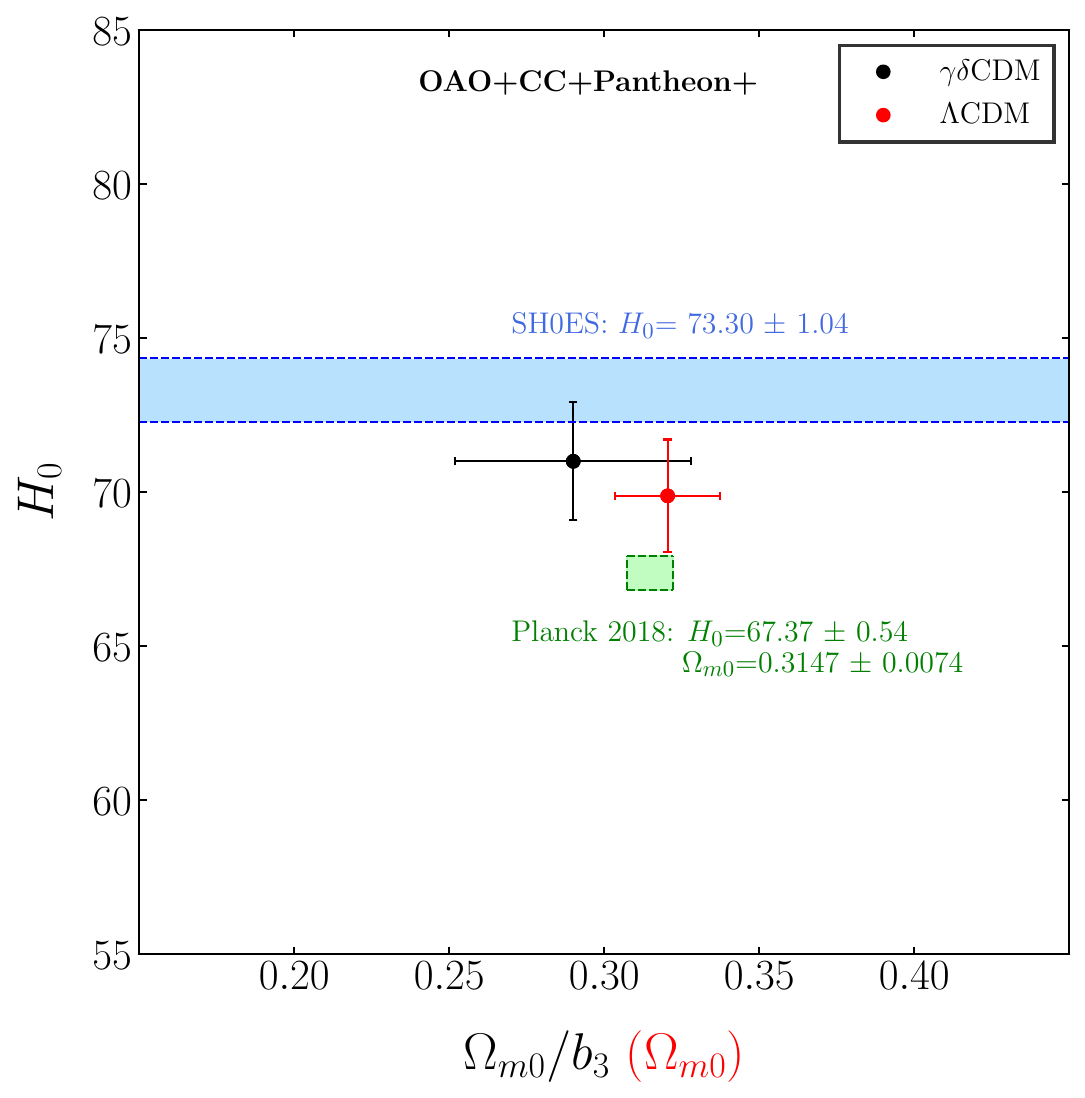}
    \label{fig:sub3b}
\end{subfigure}
\caption{The observational constraints on $H_0$ and the ``effective'' matter density, which is $\Omega_{m0}/b_3$ in $\gamma\delta$CDM model and $\Omega_{m0}$ in $\La$CDM model. Left and right panels show the median parameter values and 68\% confidence level errors obtained for the OAO+Pan+ and OAO+CC+Pan+ data sets, respectively. $H_0$ has units of $km/s/Mpc$.}
\label{Mv}
\end{figure*}

However, the value of the Hubble constant inferred from the Bayesian analysis with various data sets could not discriminate the models. In Table \ref{tmv} we note that the median and 1$\sigma$ values for Hubble constant inferred for the $\La$CDM model are in the same range as the corresponding values in the $\gamma\delta$CDM model. The main difference between two models is the value of the incubation time for the galaxies. Even though the cosmological parameters of the $\La$CDM model turns out to be near the values obtained by low-redshift observations, the incubation time in the $\La$CDM model turns out to be consistently lower than the values obtained in the $\gamma\delta$CDM model for various data sets. With the Bayesian analysis posterior values of the cosmological parameters, the $\La$CDM model fits the data with very low unrealistic incubation times $\tau$. 
As explained in the following paragraphs in detail, the population III stars are expected to form \cite{Bromm:2009uk} at about $z\sim 30$ which corresponds to the age of the Universe about $t_U \sim 100\ Myr$ in the $\La$CDM model.
Thus, the incubation time for galaxies should not be less than $100\ Myr$. 
We further tested the $\La$CDM model with SH0ES value for the Hubble constant $H_0=73.30\pm 1.04\ km/s/Mpc$ \cite{Riess:2021jrx} and fitted the matter density and astrophysical parameters. The incubation times turned out to be even lower: about $50\ Myr$ on average with the data sets containing Pantheon+ sample  \cite{Brout:2022vxf,Scolnic:2021amr}. We then tested the $\La$CDM model with Planck values for the Hubble constant $H_0=67.37\pm 0.54\ km/s/Mpc$ and the mass density $\Omega_{m0} = 0.3147\pm 0.0074$ \cite{Planck:2018vyg}. The median values for the incubation times turned out to be consistently near $110\ Myr$ in all data sets used. Such a value of the incubation time for the formation of galaxies is also in conflict with the galaxy formation models \cite{Bromm:2009uk}. In the following paragraphs we discuss this in more detail.

In Table \ref{tage} we present the age of the Universe at different redshifts in both $\ga\de$CDM and 
$\La$CDM models for various data sets, as well as in $\La$CDM model with cosmological parameters fixed by SH0ES and Planck results \cite{Riess:2021jrx,Planck:2018vyg}. To calculate the age of the Universe with SH0ES data we use $H_0=73.30\pm 1.04\ km/s/Mpc$ \cite{Riess:2021jrx} and choose $\Omega_{m0} = 0.3$. Due to a relatively high value of the Hubble constant, the age of The Universe at any redshift turns out to be smaller with this data compared to other data sets. A higher choice of $\Omega_{m0}$ \cite{Rubin:2023ovl,DES:2024tys} would make the calculated age even smaller and thus deepen the time-compression problem. For the case of the Planck data we use $H_0=67.37\pm 0.54\ km/s/Mpc$ and $\Omega_{m0} = 0.3147\pm 0.0074$ \cite{Planck:2018vyg}.

According to formation models of early galaxies, population III stars are expected to form as early as $z \sim 30$ \cite{Bromm:2009uk}, which corresponds to the age of the Universe $t_U \sim 100\ Myr$ on average in the $\La$CDM model according to various data sets. In the $\La$CDM model this value is smallest when SH0ES values is used and largest when the best fit to OAO age data is used. The corresponding age at $z = 30$ in the $\ga\de$CDM is larger by $30\ Myr$ on average in different data sets. Therefore, there is more time for the dark matter halos, and consequently the population III stars to form \cite{Bromm:2009uk} in the $\ga\de$CDM model compared to the 
$\La$CDM model. 

The incubation time for the oldest quasars is obtained by data analysis to be around $z\sim 20$ according to many studies \cite{Vagnozzi:2021tjv,Wei:2022plg,Costa:2023cmu}, which we also verified in the present work (see $z_f$ values in Table \ref{tmv}). Additionally, in Table \ref{tage} we present the age of the Universe at average $z_f = 20.5$. We find the incubation time of the oldest quasars, equal to the age of the Universe at $z_f = 20.5$, to be as small as $t_U \sim 162\ Myr$ if SH0ES values is used in the $\La$CDM model. In contrast the incubation time for the oldest quasars in the $\gamma\delta$CDM model is longer by about $45\ Myr$ on average in different data sets. Hence, there is more time for the oldest massive quasars observed by JWST to form from an initial seed of black hole of mass $M_\bullet \sim 100\ M_\odot$ \cite{Ilie:2023aqu}.

\begin{table*}[hbt!]
\centering
\def\arraystretch{1.5}
\begin{tabular}{|l|c|c|c|c|}
\hline 
\hline 
\textbf{Data Set} & \textbf{Model} & $t_U (30)$ (Myr) & $t_U (z_f)$ (Myr) & $t_U (0)$ (Gyr) \\
\hline 
SH0ES & $\Lambda$CDM & $93.06^{+1.31}_{-1.28}$
 & $161.6^{+2.29}_{-2.22}$ & $12.86^{+0.18}_{-0.18}$  \\
Planck & $\Lambda$CDM & $98.73^{+1.94}_{-1.87}$
 & $171.5^{+3.40}_{-3.28}$ & $13.80^{+0.20}_{-0.20}$  \\ 
\hline 
OAO & $\Lambda$CDM & $116.4^{+22.5}_{-24.7}$
 & $202.6^{+39.5}_{-43.2}$ & $15.68^{+2.35}_{-2.73}$ \\ 
 & $\ga\de$CDM & $147.5^{+28.7}_{-31.0}$ & $248.5^{+48.9}_{-52.5}$ & $14.82^{+2.05}_{-2.43}$ \\ 
\hline 
OAO+CC & $\Lambda$CDM & $100.0^{+15.0}_{-12.5}$
 & $173.8^{+26.2}_{-21.9}$ & $13.69^{+1.49}_{-1.30}$ \\ 
 & $\ga\de$CDM & $134.4^{+21.3}_{-19.4}$  & $227.4^{+36.4}_{-33.1}$ & $14.09^{+1.57}_{-1.47}$ \\ 
\hline 
OAO+Pan+ & $\Lambda$CDM & $92.38^{+5.93}_{-5.37}$
 & $160.4^{+10.4}_{-9.4}$ & $12.97^{+0.69}_{-0.63}$ \\ 
 & $\ga\de$CDM & $121.5^{+12.9}_{-12.1}$ & $206.9^{+22.2}_{-20.7}$ & $13.32^{+0.98}_{-0.94}$ \\ 
\hline 
OAO+CC+Pan+ & $\Lambda$CDM & $94.43^{+5.02}_{-4.71}$
 & $164.0^{+8.8}_{-8.2}$ & $13.24^{+0.56}_{-0.53}$ \\ 
 & $\ga\de$CDM & $120.5^{+11.8}_{-10.0}$ & $205.6^{+20.3}_{-17.2}$ & $13.55^{+0.87}_{-0.77}$ \\ 
\hline 
\hline
\end{tabular} 
\caption{\rm The age of the Universe with 68\% confidence level calculated in $\gamma\delta$CDM and $\Lambda$CDM models at redshifts $z=30$, $z=z_f=20.5$ and $z=0$. For the various data sets we used the posterior values of the cosmological parameters as presented in Table \ref{tmv}. We also calculate the age of the Universe with high-redshift value of $H_0=73.30\pm 1.04\ km/s/Mpc$ \cite{Riess:2021jrx} and choose $\Omega_{m0} = 0.3$, and the low-redshift value of $H_0=67.37\pm 0.54\ km/s/Mpc$ and $\Omega_{m0} = 0.3147\pm 0.0074$ \cite{Planck:2018vyg} in the $\La$CDM model. $z_f = 20.5$ is an average value for the incubation redshift of the oldest quasars (see Table \ref{tmv}).} 
\label{tage}
\end{table*}

\begin{figure*}[hbt!]
\centering
\begin{subfigure}
    \centering 
    \includegraphics[width=0.49\textwidth]{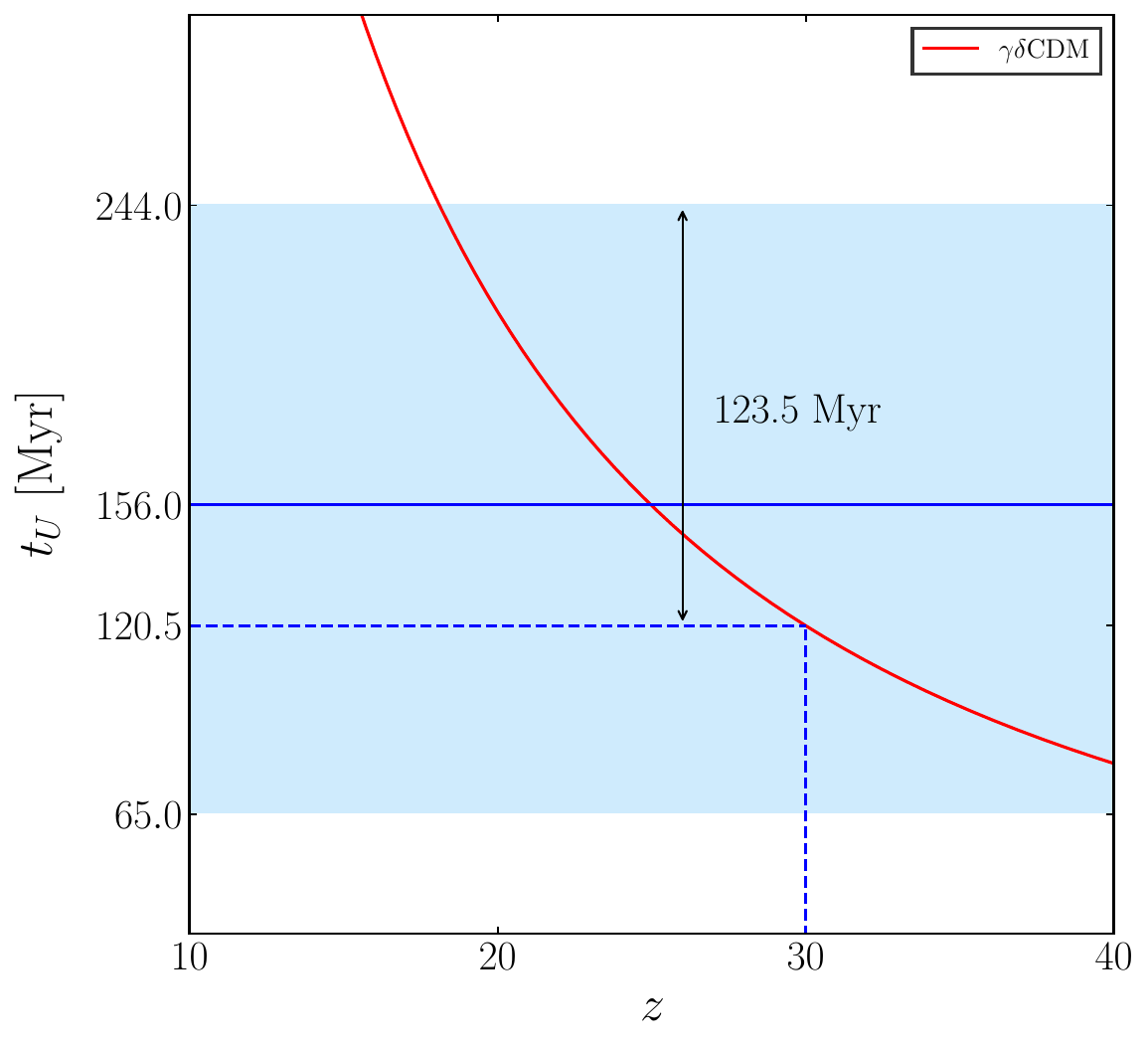}
    \label{fig:sub3a}
\end{subfigure}
\begin{subfigure}
    \centering 
    \includegraphics[width=0.49\textwidth]{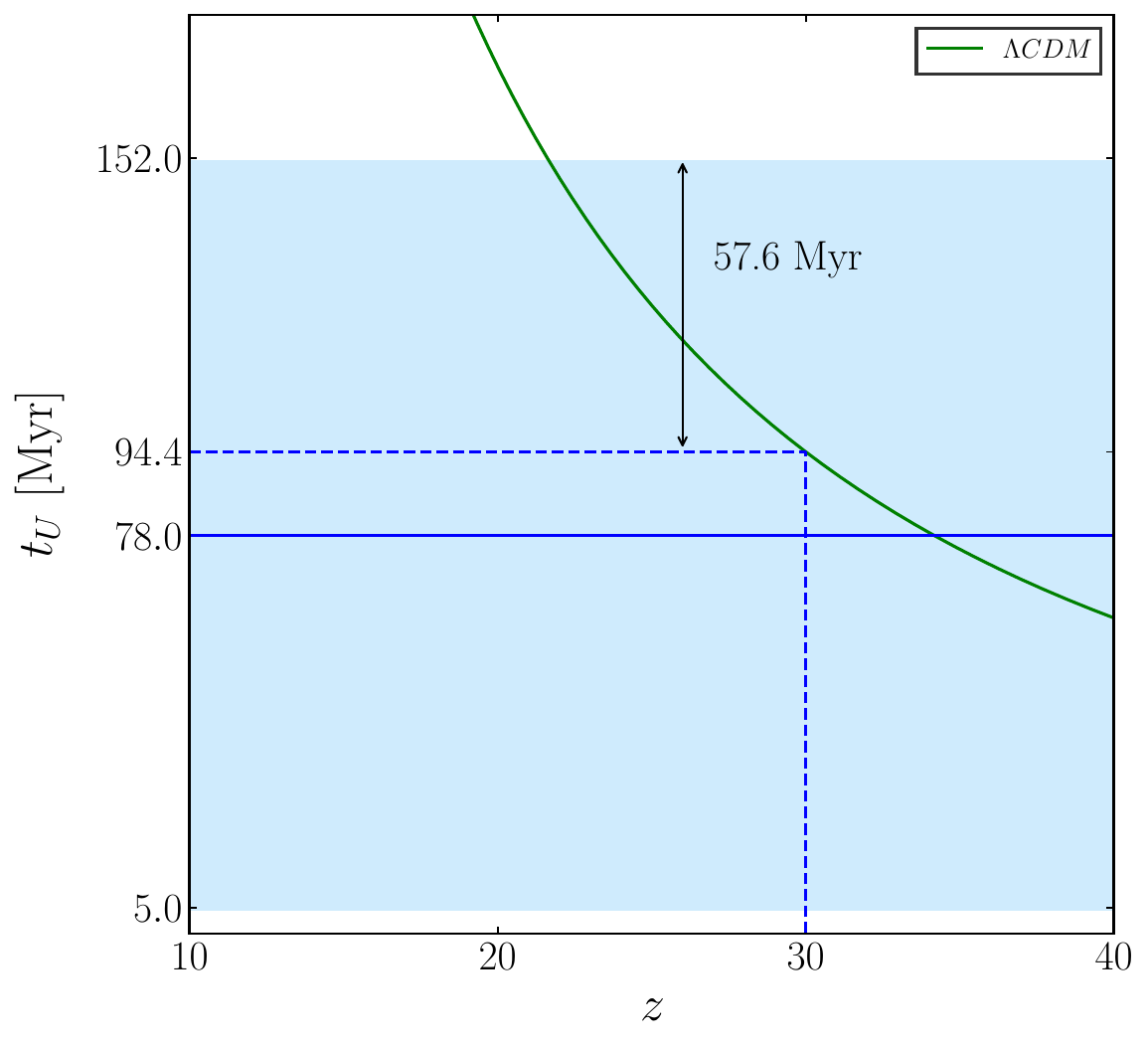}
    \label{fig:sub3b}
\end{subfigure}
\caption{Age of the Universe compared with the posterior ranges of the incubation time, $\tau$, as given in Table \ref{tmv}, in the
$\gamma\delta$CDM (left) and the $\La$CDM (right) models. Values of the cosmological parameters used to make these plots are the posterior values obtained from the Bayesian analysis with the OAO+CC+Pan+ combined data set. There is more time for the oldest galaxies and the quasars to form from initial formation of population III stars and initial seed black holes, respectively, at redshift $z\sim 30$ in the $\gamma\delta$CDM model compared to the $\La$CDM model.}
\label{ageT}
\end{figure*}

Value of the incubation time for the oldest galaxies (the parameter $\tau$ in Table \ref{tmv}) turns out not to be consistent among various data sets in the $\La$CDM model. The intersection of $1\sigma$ regions of the incubation time, $\tau$, in various data sets has the range $71.3 \lesssim \tau_{\La CDM} \lesssim 139.5\ Myr$ in the $\La$CDM model. In contrast, in the $\ga\de$CDM model the value of the incubation time for the oldest galaxies turns out to be consistent among various data sets. In this model, the intersection of $1\sigma$ regions of the incubation time, $\tau$, in various data sets has the range $80.5 \lesssim \tau_{\La CDM} \lesssim 235.2\ Myr$. We observe that in the $\La$CDM model the age of the Universe at the incubation of the earliest galaxies is in conflict with the time of formation of dark matter halos and the population III stars at $z \sim 30$ \cite{Bromm:2009uk}. However, again $\ga\de$CDM model allows more time for these galaxies to form from the initial seed of very first stars and the dark matter halos. See also Fig. \ref{ageT} for a visual comparison of two models. 

We stress that the modified time-redshift relation in the $\ga\de$CDM model allows more time for the high density structures to form in the cosmic dawn. Thus, in this model it could be possible for the quasars to grow from initial seed black holes with modest masses without need to assume super-Eddington accretion rates from its surroundings, unlike the case in $\La$CDM model \cite{Ilie:2023aqu,Inayoshi:2019fun}. Likewise, in this model there is more time allowed for the galaxies to grow starting with initial formation of dark matter halos and population III stars. We also note that in the short time allowed for the first galaxies to form in the $\La$CDM model, efficiency of converting baryons into stars turns out to be too high \cite{Forconi:2023izg,Parashari:2023cui,Wang:2023gla}.

Even though the modified time-redshift relation allows more time for the structures to form in high redshift, it does not affect the age of the Universe today \cite{Bernal:2021yli,Krishnan:2021dyb,Jimenez:2019onw} substantially. As presented in Table \ref{tage} for various data sets, at high redshift the age of the Universe is lengthened in the $\ga\de$CDM model by about $20-25\%$ with respect to $\La$CDM model, which allows more time for the structures to grow. At redshift zero, however, the age of the Universe is lengthened only by a factor of $3-4\%$. Thus, this model lengthens the age of the Universe at high redshifts where there is so called ``time-compression'' \cite{Melia:2023dsy,Boylan-Kolchin:2021fvy,Bernal:2021yli,Krishnan:2021dyb} problem, unlike Early Dark Energy models \cite{Boylan-Kolchin:2021fvy}. Along with that, this model does not unnecessarily lengthen \cite{Gupta:2023mgg} the age of the Universe today in order to resolve the time-compression problem at cosmic dawn.

Lastly, we would like to comment on the effect of the uniform prior ranges for the cosmological parameters on the marginalized posterior values of those parameters. As it is stated in Section \ref{metod}, the uniform prior range for the reduced Hubble constant in this paper ($0.55 \leqslant h \leqslant 0.85$) is narrower compared to the respective choices in \cite{Vagnozzi:2021tjv} ($0.4 \leqslant h \leqslant 1.0$) and \cite{Wei:2022plg} ($0.0 \leqslant h \leqslant 1.5$). The uniform prior range for $\Omega_{m0}$ chosen in this paper is the same as the choices made in \cite{Vagnozzi:2021tjv} and \cite{Wei:2022plg}. To assess the effect of the prior ranges on the marginalized posterior value of $H_0$ and to analyze the scientific significance of the OAO age data, we repeated the comparative Bayesian analysis for the $\gamma\delta$CDM and $\La$CDM models with the prior choices of \cite{Vagnozzi:2021tjv} and \cite{Wei:2022plg} for the Hubble constant. As it is already observed in \cite{Wei:2022plg}, the half-Gaussian likelihood favors low values for the Hubble constant. Thus, what the OAO age data constraints is actually an upper limit on $H_0$. 
In the case of uniform prior $0.55 \leqslant h \leqslant 0.85$ chosen in this work and fitting only the OAO age data, the upper 2$\sigma$ bound for the Hubble constant turns out to be $H_0 = 76.20$ in the $\gamma\delta$CDM model and $H_0 = 76.94$ in the $\La$CDM model. In the case of uniform prior $0.4 \leqslant h \leqslant 1.0$ of \cite{Vagnozzi:2021tjv} and fitting only the OAO age data, the upper 2$\sigma$ bound for the Hubble constant turns out to be $H_0 = 73.60$ in the $\gamma\delta$CDM model and $H_0 = 74.71$ in the $\La$CDM model. In the case of uniform prior $0.0 \leqslant h \leqslant 1.5$ of \cite{Wei:2022plg} and fitting again only the OAO age data, the respective values are found to be $H_0 = 71.79$ in the $\gamma\delta$CDM model and $H_0 = 74.19$ in the $\La$CDM model. In the case of the combined data sets with the either prior ranges, the marginalized posterior values for the Hubble parameter do not change significantly from the values given in the second, third and fourth columns of Table \ref{tmv}. From these analyses we conclude that the OAO age data is significant in restricting an upper limit for the Hubble constant.


\section{Conclusions \label{conc}} 

The $\La$CDM model is the standard model of cosmology, since it brings forth explanations to wide range of observational data with great success. However, various tensions that have surfaced in last few years show us that $\La$CDM model might not be the ``end of the story'' \cite{Vagnozzi:2023nrq}. The late-time and early-time modifications of $\La$CDM model have been fruitless so far to produce solutions to cosmic tensions \cite{Verde:2019ivm,Knox:2019rjx,Riess:2019qba,DiValentino:2021izs,Perivolaropoulos:2021jda,Schoneberg:2021qvd,Shah:2021onj,Abdalla:2022yfr,Vagnozzi:2023nrq}. Perhaps, simple modifications in the $\La$CDM model are not enough the resolve the various cosmic tensions, but a new approach with a new model is needed to reconcile the various observations of the various epochs of the Universe. In this motive, the initial assumptions of contemporary cosmology might need to be modified. Such modifications of the theory of cosmology comprise mostly the alternative gravity theories, in which the gravitational law of gravity is modified at cosmological scales. Even though they are founded on a modified or alternative gravity, most of these theories still construct their models based on a Friedmann equation, in which extra terms of the alternative gravity field equations are collectively interpreted as an effective energy-momentum tensor. Therefore, such models should still be considered in the paradigm of $\La$CDM model. Effective energy-momentum tensor of such theories usually bring forth modifications of the dark sectors, and thus change early- or late-time physics inside the $\La$CDM model.

In a previous work \cite{Deliduman:2023caa}, we started a study of the cosmology of an ellipsoidal Universe in the framework of $f(R)$ theory of gravity with a modified Friedmann equation. Since there are observational hints of possible anisotropy in the distribution of matter and the geometry of the Universe, as we summarized in the introduction, we modeled the background geometry of the Universe with a Bianchi type I metric. 
This distinguishes our work from other works based on $f(R)$ gravity: the Friedmann equation in our model is not in the general relativistic form. We observed in \cite{Deliduman:2023caa} that the consequences of the modified Friedmann equation (\ref{H2}) is far reaching. The contribution of the energy densities to the Universe's expansion is different than the ordinary Friedmann equation. The modification is two-fold: firstly, the ``bare'' energy densities are weighted by the coefficients $b_n$ (\ref{bn}), and secondly, redshift dependence shifts by $(1+z)^{-\delta}$ (\ref{H2}) compared to the standard $\La$CDM model \cite{Mukhanov:2005sc}. Both of these modifications depend on the value of the $\delta$ parameter (\ref{gamma}). General relativistic limit is obtained for $\delta = 0$ as expected. Existence of the $\delta$ parameter is closely related to the existence of the anisotropic stress $\Omega_{s0}$ \cite{Leach2006} in the model. 
However, this modified form of Friedmann equation do not allow existence of a cosmological constant term in the theory \cite{Deliduman:2023caa}. 
Nevertheless, we included a dark energy term with redshift dependence $(1+z)^\gamma$, with $0<\gamma<2$ that comes from the matter Lagrangian. With the addition of the shift $(1+z)^{-\delta}$, the dark energy sector in the modified Friedmann equation has redshift dependence as $(1+z)^{\gamma-\delta}$ (\ref{H2}). In \cite{Deliduman:2023caa} it is discussed whether the source of this form of dark energy could be dark energy coupled super massive black holes \cite{Farrah:2023opk,Croker:2021duf,Croker:2020plg,Gliner:1966}. 

In the present work we tested the observational relevance of the $\gamma\delta$CDM model to the age problem by constraining its parameters with the ages of oldest astronomical objects (OAO), prepared by Vagnozzi et al. \cite{Vagnozzi:2021tjv} based on galaxies and high-z quasars with redshifts up to $z\sim8$. OAO data requires a half-Gaussian for the (log-)likelihood function and thus cannot constrain the model parameters completely. Thus, we also best fit the model to cosmic chronometers (CC) Hubble data \cite{Jimenez:2001gg,Moresco:2012,Moresco:2015,Moresco:2016,Moresco:2020,Favale:2023lnp} and the Pantheon+ type Ia SNe data \cite{Brout:2022vxf,Scolnic:2021amr,Scolnic:pant} of the late Universe at low redshift. We find that, thanks to modified time--redshift relation, the $\ga\de$CDM model has more plausible time period at high redshift for the large and massive galaxies and massive quasars to form, whereas the age of the Universe today is not modified significantly.

There are many further cosmological tests to perform in order to check viability of our model. Firstly we would like to test our model with further data sets \cite{DES:2024tys,Nunes:2020hzy,eBOSS:2020yzd,deCarvalho:2021azj}. These data sets would probably constrain the cosmological parameters of the model more and show us the relevance of the $\ga\de$CDM model to explain various cosmological observations. The other important tension in the $\La$CDM model is the tension in the value of the growth parameter $S_8$ \cite{Perivolaropoulos:2021jda,Abdalla:2022yfr,DiValentino:2020vvd,Nunes:2021ipq}. We plan to analyze the value of the growth parameter in a future publication and check whether $S_8$ tension can be resolved in our model. $H_0$ and $S_8$ tensions might be connected, because as the value of $H_0$ tends to decrease with increasing redshift, the value of $S_8$ tends to decrease with decreasing redshift in the $\La$CDM model \cite{Vagnozzi:2023nrq,Esposito:2022plo,Adil:2023jtu}. We will investigate whether our model might bring an explanation to these trends.

\section*{Acknowledgments} 

Authors thank Jun-Jie Wei for providing the OAO data, and O\u{g}uzhan Ka\c{s}\i k\c{c}\i\ and Vildan Kele\c{s} Tu\u{g}yano\u{g}lu for helpful discussions. Furkan \c{S}akir Dilsiz is also supported by TUBITAK 2211/A Domestic Ph.D. scholar. The numerical calculations reported in this paper were partially performed at TUBITAK ULAKBIM, High Performance and Grid Computing Center (TRUBA resources).




\begin{thebibliography}{199}

\bibitem{Hubble:1929ig}
E.~Hubble,
``A relation between distance and radial velocity among extra-galactic nebulae,''
Proc. Nat. Acad. Sci. \textbf{15}, 168-173 (1929)
doi:10.1073/pnas.15.3.168

\bibitem{MacCallum:2015aaa}
M.~A.~H.~MacCallum,
``Milestones of general relativity: Hubble\textquoteright{}s law (1929) and the expansion of the Universe,''
Class. Quant. Grav. \textbf{32}, no.12, 124002 (2015)
doi:10.1088/0264-9381/32/12/124002
[arXiv:1504.03606 [physics.hist-ph]].

\bibitem{Freedman:2021ahq}
W.~L.~Freedman,
``Measurements of the Hubble Constant: Tensions in Perspective,''
Astrophys. J. \textbf{919}, no.1, 16 (2021)
doi:10.3847/1538-4357/ac0e95
[arXiv:2106.15656 [astro-ph.CO]].

\bibitem{Riess:2021jrx}
A.~G.~Riess, W.~Yuan, L.~M.~Macri, D.~Scolnic, D.~Brout, S.~Casertano, D.~O.~Jones, Y.~Murakami, L.~Breuval and T.~G.~Brink, \textit{et al.}
``A Comprehensive Measurement of the Local Value of the Hubble Constant with 1 km s$^{-1}$ Mpc$^{-1}$ Uncertainty from the Hubble Space Telescope and the SH0ES Team,''
Astrophys. J. Lett. \textbf{934}, no.1, L7 (2022)
doi:10.3847/2041-8213/ac5c5b
[arXiv:2112.04510 [astro-ph.CO]].

\bibitem{Planck:2018vyg}
N.~Aghanim \textit{et al.} [Planck],
``Planck 2018 results. VI. Cosmological parameters,''
Astron. Astrophys. \textbf{641}, A6 (2020)
[erratum: Astron. Astrophys. \textbf{652}, C4 (2021)]
doi:10.1051/0004-6361/201833910
[arXiv:1807.06209 [astro-ph.CO]].

\bibitem{Verde:2019ivm}
L.~Verde, T.~Treu and A.~G.~Riess,
``Tensions between the Early and the Late Universe,''
Nature Astron. \textbf{3}, 891
doi:10.1038/s41550-019-0902-0
[arXiv:1907.10625 [astro-ph.CO]].

\bibitem{Vagnozzi:2019ezj}
S.~Vagnozzi,
``New physics in light of the $H_0$ tension: An alternative view,''
Phys. Rev. D \textbf{102}, no.2, 023518 (2020)
doi:10.1103/PhysRevD.102.023518
[arXiv:1907.07569 [astro-ph.CO]].

\bibitem{Knox:2019rjx}
L.~Knox and M.~Millea,
``Hubble constant hunter\textquoteright{}s guide,''
Phys. Rev. D \textbf{101}, no.4, 043533 (2020)
doi:10.1103/PhysRevD.101.043533
[arXiv:1908.03663 [astro-ph.CO]].

\bibitem{Riess:2019qba}
A.~G.~Riess,
Nature Rev. Phys. \textbf{2}, no.1, 10-12 (2019)
doi:10.1038/s42254-019-0137-0
[arXiv:2001.03624 [astro-ph.CO]].

\bibitem{DiValentino:2021izs}
E.~Di Valentino, O.~Mena, S.~Pan, L.~Visinelli, W.~Yang, A.~Melchiorri, D.~F.~Mota, A.~G.~Riess and J.~Silk,
``In the realm of the Hubble tension\textemdash{}a review of solutions,''
Class. Quant. Grav. \textbf{38}, no.15, 153001 (2021)
doi:10.1088/1361-6382/ac086d
[arXiv:2103.01183 [astro-ph.CO]].

\bibitem{Perivolaropoulos:2021jda}
L.~Perivolaropoulos and F.~Skara,
``Challenges for \ensuremath{\Lambda}CDM: An update,''
New Astron. Rev. \textbf{95}, 101659 (2022)
doi:10.1016/j.newar.2022.101659
[arXiv:2105.05208 [astro-ph.CO]].

\bibitem{Schoneberg:2021qvd}
N.~Sch\"oneberg, G.~Franco Abell\'an, A.~P\'erez S\'anchez, S.~J.~Witte, V.~Poulin and J.~Lesgourgues,
``The H0 Olympics: A fair ranking of proposed models,''
Phys. Rept. \textbf{984}, 1-55 (2022)
doi:10.1016/j.physrep.2022.07.001
[arXiv:2107.10291 [astro-ph.CO]].

\bibitem{Shah:2021onj}
P.~Shah, P.~Lemos and O.~Lahav,
``A buyer\textquoteright{}s guide to the Hubble constant,''
Astron. Astrophys. Rev. \textbf{29}, no.1, 9 (2021)
doi:10.1007/s00159-021-00137-4
[arXiv:2109.01161 [astro-ph.CO]].

\bibitem{Abdalla:2022yfr}
E.~Abdalla, G.~Franco Abell\'an, A.~Aboubrahim, A.~Agnello, O.~Akarsu, Y.~Akrami, G.~Alestas, D.~Aloni, L.~Amendola and L.~A.~Anchordoqui, \textit{et al.}
``Cosmology intertwined: A review of the particle physics, astrophysics, and cosmology associated with the cosmological tensions and anomalies,''
JHEAp \textbf{34}, 49-211 (2022)
doi:10.1016/j.jheap.2022.04.002
[arXiv:2203.06142 [astro-ph.CO]].

\bibitem{Riess:2024ohe}
A.~G.~Riess, G.~S.~Anand, W.~Yuan, L.~M.~Macri, S.~Casertano, A.~Dolphin, L.~Breuval, D.~Scolnic, M.~Perrin and R.~I.~Anderson,
``JWST Observations Reject Unrecognized Crowding of Cepheid Photometry as an Explanation for the Hubble Tension at 8 sigma Confidence,''
[arXiv:2401.04773 [astro-ph.CO]].

\bibitem{Anand:2024nim}
G.~S.~Anand, A.~G.~Riess, W.~Yuan, R.~Beaton, S.~Casertano, S.~Li, D.~I.~Makarov, L.~N.~Makarova, R.~B.~Tully and R.~I.~Anderson, \textit{et al.}
``Tip of the Red Giant Branch Distances with JWST: An Absolute Calibration in NGC 4258 and First Applications to Type Ia Supernova Hosts,''
[arXiv:2401.04776 [astro-ph.CO]].

\bibitem{Li:2024yoe}
S.~Li, A.~G.~Riess, S.~Casertano, G.~S.~Anand, D.~M.~Scolnic, W.~Yuan, L.~Breuval and C.~D.~Huang,
``Reconnaissance with JWST of the J-region Asymptotic Giant Branch in Distance Ladder Galaxies: From Irregular Luminosity Functions to Approximation of the Hubble Constant,''
[arXiv:2401.04777 [astro-ph.CO]].

\bibitem{Freedman:2023jcz}
W.~L.~Freedman and B.~F.~Madore,
``Progress in direct measurements of the Hubble constant,''
JCAP \textbf{11}, 050 (2023)
doi:10.1088/1475-7516/2023/11/050
[arXiv:2309.05618 [astro-ph.CO]].

\bibitem{McGaugh:2023nkc}
S.~McGaugh,
``Discord in Concordance Cosmology and Anomalously Massive Early Galaxies,''
Universe \textbf{10}, 48 (2024)
doi:10.3390/Universe10010048
[arXiv:2312.03127 [astro-ph.CO]].

\bibitem{Vagnozzi:2023nrq}
S.~Vagnozzi,
``Seven Hints That Early-Time New Physics Alone Is Not Sufficient to Solve the Hubble Tension,''
Universe \textbf{9}, no.9, 393 (2023)
doi:10.3390/Universe9090393
[arXiv:2308.16628 [astro-ph.CO]].

\bibitem{Deliduman:2023caa}
C.~Deliduman, O.~Kasikci and V.~K.~Tugyanoglu,
``f(R) Gravity in an Ellipsoidal Universe,''
[arXiv:2310.02914 [gr-qc]].

\bibitem{Naidu:2022wia}
R.~P.~Naidu, P.~A.~Oesch, P.~van Dokkum, E.~J.~Nelson, K.~A.~Suess et al.,
``Two Remarkably Luminous Galaxy Candidates at z $\approx$ 10-12 Revealed by JWST,''
Astrophys. J. Lett. \textbf{940}, L14 (2022)
doi:10.3847/2041-8213/ac9b22
arXiv:2207.09434 [astro-ph.GA].

\bibitem{Castellano:2022wia}
M.~Castellano, A.~Fontana, T.~Treu, P.~Santini, E.~Merlin, et al., 
``Early Results from GLASS-JWST. III. Galaxy Candidates at z $\sim$ 9-15,''
Astrophys. J. Lett. \textbf{938}, L15 (2022)
doi:10.3847/2041-8213/ac94d0
arXiv:2207.09436 [astro-ph.GA].

\bibitem{Labbe:2022ahb}
I.~Labbe, P.~van Dokkum, E.~Nelson, R.~Bezanson, K.~A.~Suess, J.~Leja, G.~Brammer, K.~Whitaker, E.~Mathews and M.~Stefanon, \textit{et al.}
``A population of red candidate massive galaxies \textasciitilde{}600 Myr after the Big Bang,''
Nature \textbf{616}, no.7956, 266-269 (2023)
doi:10.1038/s41586-023-05786-2
[arXiv:2207.12446 [astro-ph.GA]].

\bibitem{Menci:2022wia}
N.~Menci, M.~Castellano, P.~Santini, E.~Merlin, A.~Fontana and F.~Shankar,
``High-redshift Galaxies from Early JWST Observations: Constraints on Dark Energy Models,''
Astrophys. J. Lett. \textbf{938}, no.1, L5 (2022)
doi:10.3847/2041-8213/ac96e9
[arXiv:2208.11471 [astro-ph.CO]].

\bibitem{Robertson:2022gdk}
B.~E.~Robertson, S.~Tacchella, B.~D.~Johnson, K.~Hainline, L.~Whitler, D.~J.~Eisenstein, R.~Endsley, M.~Rieke, D.~P.~Stark and S.~Alberts, \textit{et al.}
``Identification and properties of intense star-forming galaxies at redshifts z\,\ensuremath{>}\,10,''
Nature Astron. \textbf{7}, no.5, 611-621 (2023)
doi:10.1038/s41550-023-01921-1
[arXiv:2212.04480 [astro-ph.GA]].

\bibitem{Curtis-Lake:2023} 
E.~Curtis-Lake, S.~Carniani, A.~Cameron, et al.,
``Spectroscopic confirmation of four metal-poor galaxies at z=10.3-13.2''
Nature Astronomy, 7, 622 (2023)
doi:10.1038/s41550-023-01918-w
[arXiv:2212.04568 [astro-ph.GA]].

\bibitem{ArrabalHaro:2023}
P.~Arrabal Haro, M.~Dickinson, S.~L.~Finkelstein, J.S.~Kartaltepe, C.~T.~Donnan, D.~Burgarella, A.~C.~Carnall, et al., 
``Confirmation and refutation of very luminous galaxies in the early Universe,''
Nature {\bf 622}, 707 (2023) 
doi:10.1038/s41586-023-06521-7
[arXiv:2303.15431 [astro-ph.GA]].

\bibitem{Yung:2023bng}
L.~Y.~A.~Yung, R.~S.~Somerville, S.~L.~Finkelstein, S.~M.~Wilkins and J.~P.~Gardner,
``Are the ultra-high-redshift galaxies at z \ensuremath{>} 10 surprising in the context of standard galaxy formation models?,''
doi:10.1093/mnras/stad3484
[arXiv:2304.04348 [astro-ph.GA]].

\bibitem{Forconi:2023izg}
M.~Forconi, Ruchika, A.~Melchiorri, O.~Mena and N.~Menci,
``Do the early galaxies observed by JWST disagree with Planck's CMB polarization measurements?,''
JCAP \textbf{10}, 012 (2023)
doi:10.1088/1475-7516/2023/10/012
[arXiv:2306.07781 [astro-ph.CO]].

\bibitem{Maiolino:2023zdu}
R.~Maiolino, J.~Scholtz, J.~Witstok, S.~Carniani, F.~D'Eugenio, A.~de Graaff, H.~Uebler, S.~Tacchella, E.~Curtis-Lake and S.~Arribas, \textit{et al.}
``A small and vigorous black hole in the early Universe,''
[arXiv:2305.12492 [astro-ph.GA]].

\bibitem{Bogdan:2023ilu}
A.~Bogdan, A.~D.~Goulding, P.~Natarajan, O.~E.~Kov\'acs, G.~R.~Tremblay, U.~Chadayammuri, M.~Volonteri, R.~P.~Kraft, W.~R.~Forman and C.~Jones, \textit{et al.}
``Evidence for heavy-seed origin of early supermassive black holes from a z\,\ensuremath{\approx}\,10 X-ray quasar,''
Nature Astron. \textbf{8}, no.1, 126-133 (2024)
doi:10.1038/s41550-023-02111-9
[arXiv:2305.15458 [astro-ph.GA]].

\bibitem{Maiolino:2023bpi}
R.~Maiolino, J.~Scholtz, E.~Curtis-Lake, S.~Carniani, W.~Baker, A.~de Graaff, S.~Tacchella, H.~\"Ubler, F.~D'Eugenio and J.~Witstok, \textit{et al.}
``JADES. The diverse population of infant Black Holes at 4\ensuremath{<}z\ensuremath{<}11: merging, tiny, poor, but mighty,''
[arXiv:2308.01230 [astro-ph.GA]].

\bibitem{Natarajan:2023rxq}
P.~Natarajan, F.~Pacucci, A.~Ricarte, A.~Bogdan, A.~D.~Goulding and N.~Cappelluti,
``First Detection of an Overmassive Black Hole Galaxy UHZ1: Evidence for Heavy Black Hole Seed Formation from Direct Collapse,''
Astrophys. J. Lett. \textbf{960}, no.1, L1 (2024)
doi:10.3847/2041-8213/ad0e76
[arXiv:2308.02654 [astro-ph.HE]].

\bibitem{Furtak:2023ege}
L.~J.~Furtak, I.~Labb\'e, A.~Zitrin, J.~E.~Greene, P.~Dayal, I.~Chemerynska, V.~Kokorev, T.~B.~Miller, A.~D.~Goulding and R.~Bezanson, \textit{et al.}
``A supermassive black hole in the early Universe growing in the shadows,''
[arXiv:2308.05735 [astro-ph.GA]].

\bibitem{Pacucci:2023oci}
F.~Pacucci, B.~Nguyen, S.~Carniani, R.~Maiolino and X.~Fan,
``JWST CEERS and JADES Active Galaxies at z = 4\textendash{}7 Violate the Local M $_{?}$\textendash{}M $_{?}$ Relation at \ensuremath{>}3\ensuremath{\sigma}: Implications for Low-mass Black Holes and Seeding Models,''
Astrophys. J. Lett. \textbf{957}, no.1, L3 (2023)
doi:10.3847/2041-8213/ad0158
[arXiv:2308.12331 [astro-ph.GA]].

\bibitem{Ilie:2023aqu}
C.~Ilie, K.~Freese, A.~Petric and J.~Paulin,
``UHZ1 and the other three most distant quasars observed: possible evidence for Supermassive Dark Stars,''
[arXiv:2312.13837 [astro-ph.GA]].

\bibitem{Boylan-Kolchin:2022kae}
M.~Boylan-Kolchin,
``Stress testing \ensuremath{\Lambda}CDM with high-redshift galaxy candidates,''
Nature Astron. \textbf{7}, no.6, 731-735 (2023)
doi:10.1038/s41550-023-01937-7
[arXiv:2208.01611 [astro-ph.CO]].

\bibitem{Lovell:2022bhx}
C.~C.~Lovell, I.~Harrison, Y.~Harikane, S.~Tacchella and S.~M.~Wilkins,
``Extreme value statistics of the halo and stellar mass distributions at high redshift: are JWST results in tension with \ensuremath{\Lambda}CDM?,''
Mon. Not. Roy. Astron. Soc. \textbf{518}, no.2, 2511-2520 (2022)
doi:10.1093/mnras/stac3224
[arXiv:2208.10479 [astro-ph.GA]].

\bibitem{Gimenez-Arteaga:2022ubw}
C.~Gim\'enez-Arteaga, P.~A.~Oesch, G.~B.~Brammer, F.~Valentino, C.~A.~Mason, A.~Weibel, L.~Barrufet, S.~Fujimoto, K.~E.~Heintz and E.~J.~Nelson, \textit{et al.}
``Spatially Resolved Properties of Galaxies at 5 \ensuremath{<} z \ensuremath{<} 9 in the SMACS 0723 JWST ERO Field,''
Astrophys. J. \textbf{948}, no.2, 126 (2023)
doi:10.3847/1538-4357/acc5ea
[arXiv:2212.08670 [astro-ph.GA]].

\bibitem{Cameron:2023zhm}
A.~J.~Cameron, H.~Katz, M.~P.~Rey and A.~Saxena,
``Nitrogen enhancements 440\,Myr after the big bang: supersolar N/O, a tidal disruption event, or a dense stellar cluster in GN-z11?,''
Mon. Not. Roy. Astron. Soc. \textbf{523}, no.3, 3516-3525 (2023)
doi:10.1093/mnras/stad1579
[arXiv:2302.10142 [astro-ph.GA]].

\bibitem{Melia:2023dsy}
F.~Melia,
``The cosmic timeline implied by the JWST high-redshift galaxies,''
Mon. Not. Roy. Astron. Soc. \textbf{521}, no.1, L85-L89 (2023)
doi:10.1093/mnrasl/slad025
[arXiv:2302.10103 [astro-ph.CO]].

\bibitem{Parashari:2023cui}
P.~Parashari and R.~Laha,
``Primordial power spectrum in light of JWST observations of high redshift galaxies,''
Mon. Not. Roy. Astron. Soc. \textbf{526}, no.1, L63-L69 (2023)
doi:10.1093/mnrasl/slad107
[arXiv:2305.00999 [astro-ph.CO]].

\bibitem{Hirano:2023auh}
S.~Hirano and N.~Yoshida,
``Early Structure Formation from Primordial Density Fluctuations with a Blue, Tilted Power Spectrum: High-Redshift Galaxies,''
[arXiv:2306.11993 [astro-ph.GA]].

\bibitem{Adil:2023ara}
S.~A.~Adil, U.~Mukhopadhyay, A.~A.~Sen and S.~Vagnozzi,
``Dark energy in light of the early JWST observations: case for a negative cosmological constant?,''
JCAP \textbf{10}, 072 (2023)
doi:10.1088/1475-7516/2023/10/072
[arXiv:2307.12763 [astro-ph.CO]].

\bibitem{Menci:2024rbq}
N.~Menci, S.~A.~Adil, U.~Mukhopadhyay, A.~A.~Sen and S.~Vagnozzi,
``Negative cosmological constant in the dark energy sector: tests from JWST photometric and spectroscopic observations of high-redshift galaxies,''
[arXiv:2401.12659 [astro-ph.CO]].

\bibitem{Sun:2023wqq}
C.~Sun, M.~A.~Buen-Abad and J.~Fan,
``Probing New Physics with High-Redshift Quasars: Axions and Non-standard Cosmology,''
[arXiv:2309.07212 [astro-ph.CO]].

\bibitem{vanPutten:2023ths}
M.~H.~P.~M.~van Putten,
``The Fast and Furious in JWST high-z galaxies,''
Phys. Dark Univ. \textbf{43}, 101417 (2024)
doi:10.1016/j.dark.2023.101417
[arXiv:2312.16692 [astro-ph.CO]].

\bibitem{Forconi:2023hsj}
M.~Forconi, W.~Giar\`e, O.~Mena, Ruchika, E.~Di Valentino, A.~Melchiorri and R.~C.~Nunes,
``A double take on early and interacting dark energy from JWST,''
[arXiv:2312.11074 [astro-ph.CO]].

\bibitem{Laursen:2023tbg}
P.~Laursen,
``Galaxy formation from a timescale perspective,''
doi:10.1007/16618\_2023\_71
[arXiv:2309.02486 [astro-ph.GA]].

\bibitem{Davari:2023tam}
Z.~Davari, A.~Ashoorioon and K.~Rezazadeh,
``Spherical Collapse Approach for Non-standard Cold Dark Matter Models and Enhanced Early Galaxy Formation in JWST,''
[arXiv:2311.15083 [astro-ph.CO]].

\bibitem{Melia:2014cva}
F.~Melia,
``The Premature Formation of High Redshift Galaxies,''
Astron. J. \textbf{147}, 120 (2014)
doi:10.1088/0004-6256/147/5/120
[arXiv:1403.0908 [astro-ph.GA]].

\bibitem{Salvaterra:2010nb}
R.~Salvaterra, A.~Ferrara and P.~Dayal,
``Simulating high-redshift galaxies,''
Mon. Not. Roy. Astron. Soc. \textbf{414}, 847 (2011)
doi:10.1111/j.1365-2966.2010.18155.x
[arXiv:1003.3873 [astro-ph.CO]].

\bibitem{Jaacks:2012rn}
J.~Jaacks, K.~Nagamine and J.~H.~Choi,
``Duty Cycle and the Increasing Star Formation History of z\ensuremath{>}=6 Galaxies,''
Mon. Not. Roy. Astron. Soc. \textbf{427}, 403 (2012)
doi:10.1111/j.1365-2966.2012.21989.x
[arXiv:1204.4846 [astro-ph.CO]].

\bibitem{Boylan-Kolchin:2021fvy}
M.~Boylan-Kolchin and D.~R.~Weisz,
``Uncertain times: the redshift\textendash{}time relation from cosmology and stars,''
Mon. Not. Roy. Astron. Soc. \textbf{505}, no.2, 2764-2783 (2021)
doi:10.1093/mnras/stab1521
[arXiv:2103.15825 [astro-ph.CO]].

\bibitem{Padmanabhan:2023esp}
H.~Padmanabhan and A.~Loeb,
``Alleviating the Need for Exponential Evolution of JWST Galaxies in 10$^{10}$ M $_{?}$ Haloes at z \ensuremath{>} 10 by a Modified \ensuremath{\Lambda}CDM Power Spectrum,''
Astrophys. J. Lett. \textbf{953}, no.1, L4 (2023)
doi:10.3847/2041-8213/acea7a
[arXiv:2306.04684 [astro-ph.CO]].

\bibitem{Inayoshi:2019fun}
K.~Inayoshi, E.~Visbal and Z.~Haiman,
``The Assembly of the First Massive Black Holes,''
Ann. Rev. Astron. Astrophys. \textbf{58}, 27-97 (2020)
doi:10.1146/annurev-astro-120419-014455
[arXiv:1911.05791 [astro-ph.GA]].

\bibitem{Bernal:2021yli}
J.~L.~Bernal, L.~Verde, R.~Jimenez, M.~Kamionkowski, D.~Valcin and B.~D.~Wandelt,
``The trouble beyond $H_0$ and the new cosmic triangles,''
Phys. Rev. D \textbf{103}, no.10, 103533 (2021)
doi:10.1103/PhysRevD.103.103533
[arXiv:2102.05066 [astro-ph.CO]].

\bibitem{Krishnan:2021dyb}
C.~Krishnan, R.~Mohayaee, E.~\'O.~Colg\'ain, M.~M.~Sheikh-Jabbari and L.~Yin,
``Does Hubble tension signal a breakdown in FLRW cosmology?,''
Class. Quant. Grav. \textbf{38}, no.18, 184001 (2021)
doi:10.1088/1361-6382/ac1a81
[arXiv:2105.09790 [astro-ph.CO]].

\bibitem{Gupta:2023mgg}
R.~P.~Gupta,
``JWST early Universe observations and \ensuremath{\Lambda}CDM cosmology,''
Mon. Not. Roy. Astron. Soc. \textbf{524}, no.3, 3385-3395 (2023)
doi:10.1093/mnras/stad2032
[arXiv:2309.13100 [astro-ph.CO]].

\bibitem{Ying:2023oie}
J.~Ying, B.~Chaboyer, E.~M.~Boudreaux, C.~Slaughter, M.~Boylan-Kolchin and D.~Weisz,
``The Absolute Age of M92,''
Astron. J. \textbf{166}, no.1, 18 (2023)
doi:10.3847/1538-3881/acd9b1
[arXiv:2306.02180 [astro-ph.SR]].

\bibitem{Sotiriou:2008rp}
T.~P.~Sotiriou and V.~Faraoni,
``f(R) Theories Of Gravity,''
Rev. Mod. Phys. \textbf{82}, 451-497 (2010)
doi:10.1103/RevModPhys.82.451
[arXiv:0805.1726 [gr-qc]].

\bibitem{DeFelice:2010aj}
A.~De Felice and S.~Tsujikawa,
``f(R) theories,''
Living Rev. Rel. \textbf{13}, 3 (2010)
doi:10.12942/lrr-2010-3
[arXiv:1002.4928 [gr-qc]].

\bibitem{Nojiri:2010wj}
S.~Nojiri and S.~D.~Odintsov,
``Unified cosmic history in modified gravity: from F(R) theory to Lorentz non-invariant models,''
Phys. Rept. \textbf{505}, 59-144 (2011)
doi:10.1016/j.physrep.2011.04.001
[arXiv:1011.0544 [gr-qc]].

\bibitem{Nojiri:2017ncd}
S.~Nojiri, S.~D.~Odintsov and V.~K.~Oikonomou,
``Modified Gravity Theories on a Nutshell: Inflation, Bounce and Late-time Evolution,''
Phys. Rept. \textbf{692}, 1-104 (2017)
doi:10.1016/j.physrep.2017.06.001
[arXiv:1705.11098 [gr-qc]].

\bibitem{COBE:1992syq}
G.~F.~Smoot \textit{et al.} [COBE],
``Structure in the COBE differential microwave radiometer first year maps,''
Astrophys. J. Lett. \textbf{396}, L1-L5 (1992)
doi:10.1086/186504

\bibitem{Kogut:1996us}
A.~Kogut, G.~Hinshaw, A.~J.~Banday, C.~L.~Bennett, K.~Gorski, G.~F.~Smoot and E.~L.~Wright,
``Microwave emission at high Galactic latitudes,''
Astrophys. J. Lett. \textbf{464}, L5-L9 (1996)
doi:10.1086/310072
[arXiv:astro-ph/9601060 [astro-ph]].

\bibitem{Bennett:2010jb}
C.~L.~Bennett, R.~S.~Hill, G.~Hinshaw, D.~Larson, K.~M.~Smith, J.~Dunkley, B.~Gold, M.~Halpern, N.~Jarosik and A.~Kogut, \textit{et al.}
``Seven-Year Wilkinson Microwave Anisotropy Probe (WMAP) Observations: Are There Cosmic Microwave Background Anomalies?,''
Astrophys. J. Suppl. \textbf{192}, 17 (2011)
doi:10.1088/0067-0049/192/2/17
[arXiv:1001.4758 [astro-ph.CO]].

\bibitem{deOliveira-Costa:2003utu}
A.~de Oliveira-Costa, M.~Tegmark, M.~Zaldarriaga and A.~Hamilton,
``The Significance of the largest scale CMB fluctuations in WMAP,''
Phys. Rev. D \textbf{69}, 063516 (2004)
doi:10.1103/PhysRevD.69.063516
[arXiv:astro-ph/0307282 [astro-ph]].

\bibitem{WMAP:2003ivt}
C.~L.~Bennett \textit{et al.} [WMAP],
``First year Wilkinson Microwave Anisotropy Probe (WMAP) observations: Preliminary maps and basic results,''
Astrophys. J. Suppl. \textbf{148}, 1-27 (2003)
doi:10.1086/377253
[arXiv:astro-ph/0302207 [astro-ph]].

\bibitem{WMAP:2003zzr}
G.~Hinshaw \textit{et al.} [WMAP],
``First year Wilkinson Microwave Anisotropy Probe (WMAP) observations: The Angular power spectrum,''
Astrophys. J. Suppl. \textbf{148}, 135 (2003)
doi:10.1086/377225
[arXiv:astro-ph/0302217 [astro-ph]].

\bibitem{WMAP:2003elm}
D.~N.~Spergel \textit{et al.} [WMAP],
``First year Wilkinson Microwave Anisotropy Probe (WMAP) observations: Determination of cosmological parameters,''
Astrophys. J. Suppl. \textbf{148}, 175-194 (2003)
doi:10.1086/377226
[arXiv:astro-ph/0302209 [astro-ph]].

\bibitem{Planck:2015igc}
P.~A.~R.~Ade \textit{et al.} [Planck],
``Planck 2015 results. XVI. Isotropy and statistics of the CMB,''
Astron. Astrophys. \textbf{594}, A16 (2016)
doi:10.1051/0004-6361/201526681
[arXiv:1506.07135 [astro-ph.CO]].

\bibitem{Planck:2019kim}
Y.~Akrami \textit{et al.} [Planck],
``Planck 2018 results. IX. Constraints on primordial non-Gaussianity,''
Astron. Astrophys. \textbf{641}, A9 (2020)
doi:10.1051/0004-6361/201935891
[arXiv:1905.05697 [astro-ph.CO]].

\bibitem{Buchert:2015wwr}
T.~Buchert, A.~A.~Coley, H.~Kleinert, B.~F.~Roukema and D.~L.~Wiltshire,
``Observational Challenges for the Standard FLRW Model,''
Int. J. Mod. Phys. D \textbf{25}, no.03, 1630007 (2016)
doi:10.1142/S021827181630007X
[arXiv:1512.03313 [astro-ph.CO]].

\bibitem{Schwarz:2015cma}
D.~J.~Schwarz, C.~J.~Copi, D.~Huterer and G.~D.~Starkman,
``CMB Anomalies after Planck,''
Class. Quant. Grav. \textbf{33}, no.18, 184001 (2016)
doi:10.1088/0264-9381/33/18/184001
[arXiv:1510.07929 [astro-ph.CO]].

\bibitem{Campanelli:2006vb}
L.~Campanelli, P.~Cea and L.~Tedesco,
``Ellipsoidal Universe Can Solve The CMB Quadrupole Problem,''
Phys. Rev. Lett. \textbf{97}, 131302 (2006)
[erratum: Phys. Rev. Lett. \textbf{97}, 209903 (2006)]
doi:10.1103/PhysRevLett.97.131302
[arXiv:astro-ph/0606266 [astro-ph]].

\bibitem{Cea:2022mtf}
P.~Cea,
``The Ellipsoidal Universe and the Hubble tension,''
[arXiv:2201.04548 [astro-ph.CO]].

\bibitem{Campanelli:2007qn}
L.~Campanelli, P.~Cea and L.~Tedesco,
``Cosmic Microwave Background Quadrupole and Ellipsoidal Universe,''
Phys. Rev. D \textbf{76}, 063007 (2007)
doi:10.1103/PhysRevD.76.063007
[arXiv:0706.3802 [astro-ph]].

\bibitem{Cea:2019gnu}
P.~Cea,
``Confronting the Ellipsoidal Universe to the Planck 2018 Data,''
Eur. Phys. J. Plus \textbf{135}, no.2, 150 (2020)
doi:10.1140/epjp/s13360-020-00166-5
[arXiv:1909.05111 [astro-ph.CO]].

\bibitem{Copi:2013jna}
C.~J.~Copi, D.~Huterer, D.~J.~Schwarz and G.~D.~Starkman,
``Large-scale alignments from WMAP and Planck,''
Mon. Not. Roy. Astron. Soc. \textbf{449}, no.4, 3458-3470 (2015)
doi:10.1093/mnras/stv501
[arXiv:1311.4562 [astro-ph.CO]].

\bibitem{Land:2005ad}
K.~Land and J.~Magueijo,
``The Axis of evil,''
Phys. Rev. Lett. \textbf{95}, 071301 (2005)
doi:10.1103/PhysRevLett.95.071301
[arXiv:astro-ph/0502237 [astro-ph]].

\bibitem{Schwarz:2004gk}
D.~J.~Schwarz, G.~D.~Starkman, D.~Huterer and C.~J.~Copi,
``Is the low-l microwave background cosmic?,''
Phys. Rev. Lett. \textbf{93}, 221301 (2004)
doi:10.1103/PhysRevLett.93.221301
[arXiv:astro-ph/0403353 [astro-ph]].

\bibitem{Planck:2019evm}
Y.~Akrami \textit{et al.} [Planck],
``Planck 2018 results. VII. Isotropy and Statistics of the CMB,''
Astron. Astrophys. \textbf{641}, A7 (2020)
doi:10.1051/0004-6361/201935201
[arXiv:1906.02552 [astro-ph.CO]].

\bibitem{Mukherjee:2015mma}
S.~Mukherjee, P.~K.~Aluri, S.~Das, S.~Shaikh and T.~Souradeep,
``Direction dependence of cosmological parameters due to cosmic hemispherical asymmetry,''
JCAP \textbf{06}, 042 (2016)
doi:10.1088/1475-7516/2016/06/042
[arXiv:1510.00154 [astro-ph.CO]].

\bibitem{Javanmardi:2016whx}
B.~Javanmardi and P.~Kroupa,
``Anisotropy in the all-sky distribution of galaxy morphological types,''
Astron. Astrophys. \textbf{597}, A120 (2017)
doi:10.1051/0004-6361/201629408
[arXiv:1609.06719 [astro-ph.GA]].

\bibitem{Axelsson:2013mva}
M.~Axelsson, Y.~Fantaye, F.~K.~Hansen, A.~J.~Banday, H.~K.~Eriksen and K.~M.~Gorski,
``Directional dependence of $\Lambda$CDM cosmological parameters,''
Astrophys. J. Lett. \textbf{773}, L3 (2013)
doi:10.1088/2041-8205/773/1/L3
[arXiv:1303.5371 [astro-ph.CO]].

\bibitem{Eriksen:2007pc}
H.~K.~Eriksen, A.~J.~Banday, K.~M.~Gorski, F.~K.~Hansen and P.~B.~Lilje,
``Hemispherical power asymmetry in the three-year Wilkinson Microwave Anisotropy Probe sky maps,''
Astrophys. J. Lett. \textbf{660}, L81-L84 (2007)
doi:10.1086/518091
[arXiv:astro-ph/0701089 [astro-ph]].

\bibitem{Vielva:2003et}
P.~Vielva, E.~Martinez-Gonzalez, R.~B.~Barreiro, J.~L.~Sanz and L.~Cayon,
``Detection of non-Gaussianity in the WMAP 1 - year data using spherical wavelets,''
Astrophys. J. \textbf{609}, 22-34 (2004)
doi:10.1086/421007
[arXiv:astro-ph/0310273 [astro-ph]].

\bibitem{Cruz:2006sv}
M.~Cruz, M.~Tucci, E.~Martinez-Gonzalez and P.~Vielva,
``The non-gaussian cold spot in wmap: significance, morphology and foreground contribution,''
Mon. Not. Roy. Astron. Soc. \textbf{369}, 57-67 (2006)
doi:10.1111/j.1365-2966.2006.10312.x
[arXiv:astro-ph/0601427 [astro-ph]].

\bibitem{Luongo:2021nqh}
O.~Luongo, M.~Muccino, E.~\'O.~Colg\'ain, M.~M.~Sheikh-Jabbari and L.~Yin,
``Larger H0 values in the CMB dipole direction,''
Phys. Rev. D \textbf{105}, no.10, 103510 (2022)
doi:10.1103/PhysRevD.105.103510
[arXiv:2108.13228 [astro-ph.CO]].

\bibitem{Krishnan:2021jmh}
C.~Krishnan, R.~Mohayaee, E.~\'O.~Colg\'ain, M.~M.~Sheikh-Jabbari and L.~Yin,
``Hints of FLRW breakdown from supernovae,''
Phys. Rev. D \textbf{105}, no.6, 063514 (2022)
doi:10.1103/PhysRevD.105.063514
[arXiv:2106.02532 [astro-ph.CO]].

\bibitem{Rodrigues:2007ny}
D.~C.~Rodrigues,
``Anisotropic Cosmological Constant and the CMB Quadrupole Anomaly,''
Phys. Rev. D \textbf{77}, 023534 (2008)
doi:10.1103/PhysRevD.77.023534
[arXiv:0708.1168 [astro-ph]].

\bibitem{Bridges:2007ne}
M.~Bridges, J.~D.~McEwen, M.~Cruz, M.~P.~Hobson, A.~N.~Lasenby, P.~Vielva and E.~Martinez-Gonzalez,
``Bianchi VII\_h models and the cold spot texture,''
Mon. Not. Roy. Astron. Soc. \textbf{390}, 1372 (2008)
doi:10.1111/j.1365-2966.2008.13835.x
[arXiv:0712.1789 [astro-ph]].

\bibitem{Migkas:2020fza}
K.~Migkas, G.~Schellenberger, T.~H.~Reiprich, F.~Pacaud, M.~E.~Ramos-Ceja and L.~Lovisari,
``Probing cosmic isotropy with a new X-ray galaxy cluster sample through the $L_{\text{X}}-T$ scaling relation,''
Astron. Astrophys. \textbf{636}, A15 (2020)
doi:10.1051/0004-6361/201936602
[arXiv:2004.03305 [astro-ph.CO]].

\bibitem{Aluri:2022hzs}
P.~K.~Aluri, P.~Cea, P.~Chingangbam, M.~C.~Chu, R.~G.~Clowes, D.~Hutsem\'ekers, J.~P.~Kochappan, A.~M.~Lopez, L.~Liu and N.~C.~M.~Martens, \textit{et al.}
``Is the observable Universe consistent with the cosmological principle?,''
Class. Quant. Grav. \textbf{40}, no.9, 094001 (2023)
doi:10.1088/1361-6382/acbefc
[arXiv:2207.05765 [astro-ph.CO]].

\bibitem{Collins:1973lda}
C.~B.~Collins and S.~W.~Hawking,
``The rotation and distortion of the Universe,''
Mon. Not. Roy. Astron. Soc. \textbf{162}, 307-320 (1973)

\bibitem{Akarsu:2019pwn}
\"O.~Akarsu, S.~Kumar, S.~Sharma and L.~Tedesco,
``Constraints on a Bianchi type I spacetime extension of the standard $\Lambda$CDM model,''
Phys. Rev. D \textbf{100}, no.2, 023532 (2019)
doi:10.1103/PhysRevD.100.023532
[arXiv:1905.06949 [astro-ph.CO]].

\bibitem{Akarsu:2020pka}
\"O.~Akarsu, N.~Kat\i rc\i , A.~A.~Sen and J.~A.~Vazquez,
``Scalar field emulator via anisotropically deformed vacuum energy: Application to dark energy,''
[arXiv:2004.14863 [gr-qc]].

\bibitem{Akarsu:2021max}
\"O.~Akarsu, E.~Di Valentino, S.~Kumar, M.~Ozyigit and S.~Sharma,
``Testing spatial curvature and anisotropic expansion on top of the \ensuremath{\Lambda}CDM model,''
Phys. Dark Univ. \textbf{39}, 101162 (2023)
doi:10.1016/j.dark.2022.101162
[arXiv:2112.07807 [astro-ph.CO]].

\bibitem{Tedesco:2018dbn}
L.~Tedesco,
``Ellipsoidal Expansion of the Universe, Cosmic Shear, Acceleration and Jerk Parameter,''
Eur. Phys. J. Plus \textbf{133}, no.5, 188 (2018)
doi:10.1140/epjp/i2018-12034-x
[arXiv:1804.11203 [gr-qc]].

\bibitem{Amirhashchi:2018bic}
H.~Amirhashchi and S.~Amirhashchi,
``Current Constraints on Anisotropic and Isotropic Dark Energy Models,''
Phys. Rev. D \textbf{99}, no.2, 023516 (2019)
doi:10.1103/PhysRevD.99.023516
[arXiv:1803.08447 [astro-ph.CO]].

\bibitem{Hossienkhani:2014zoa}
H.~Hossienkhani and A.~Pasqua,
``Thermal relic abundance and anisotropy due to modified gravity,''
Astrophys. Space Sci. \textbf{349}, 39-47 (2014)
doi:10.1007/s10509-013-1645-5

\bibitem{Nojiri:2022idp}
S.~Nojiri, S.~D.~Odintsov, V.~K.~Oikonomou and A.~Constantini,
``Formalizing anisotropic inflation in modified gravity,''
Nucl. Phys. B \textbf{985}, 116011 (2022)
doi:10.1016/j.nuclphysb.2022.116011
[arXiv:2210.16383 [gr-qc]].

\bibitem{Barrow:2005qv}
J.~D.~Barrow and S.~Hervik,
``Anisotropically inflating Universes,''
Phys. Rev. D \textbf{73}, 023007 (2006)
doi:10.1103/PhysRevD.73.023007
[arXiv:gr-qc/0511127 [gr-qc]].

\bibitem{Starobinsky:1962}
A.~A.~Starobinsky,
``Isotropization of arbitrary cosmological expansion given an effective cosmological constant,''
JETP Lett. \textbf{37}, 86 (1983).

\bibitem{Wald:1983ky}
R.~M.~Wald,
``Asymptotic behavior of homogeneous cosmological models in the presence of a positive cosmological constant,''
Phys. Rev. D \textbf{28}, 2118-2120 (1983)
doi:10.1103/PhysRevD.28.2118

\bibitem{Capozziello:2005mj}
S.~Capozziello, S.~Nojiri and S.~D.~Odintsov,
``Dark energy: The Equation of state description versus scalar-tensor or modified gravity,''
Phys. Lett. B \textbf{634}, 93-100 (2006)
doi:10.1016/j.physletb.2006.01.065
[arXiv:hep-th/0512118 [hep-th]].

\bibitem{Nojiri:2006ri}
S.~Nojiri and S.~D.~Odintsov,
``Introduction to modified gravity and gravitational alternative for dark energy,''
eConf \textbf{C0602061}, 06 (2006)
doi:10.1142/S0219887807001928
[arXiv:hep-th/0601213 [hep-th]].

\bibitem{Capozziello:2006dj}
S.~Capozziello, S.~Nojiri, S.~D.~Odintsov and A.~Troisi,
``Cosmological viability of f(R)-gravity as an ideal fluid and its compatibility with a matter dominated phase,''
Phys. Lett. B \textbf{639}, 135-143 (2006)
doi:10.1016/j.physletb.2006.06.034
[arXiv:astro-ph/0604431 [astro-ph]].

\bibitem{Faraoni:2018qdr}
V.~Faraoni and J.~Cot\'e,
``Imperfect fluid description of modified gravities,''
Phys. Rev. D \textbf{98}, no.8, 084019 (2018)
doi:10.1103/PhysRevD.98.084019
[arXiv:1808.02427 [gr-qc]].

\bibitem{Vagnozzi:2021tjv}
S.~Vagnozzi, F.~Pacucci and A.~Loeb,
``Implications for the Hubble tension from the ages of the oldest astrophysical objects,''
JHEAp \textbf{36}, 27-35 (2022)
doi:10.1016/j.jheap.2022.07.004
[arXiv:2105.10421 [astro-ph.CO]].

\bibitem{Jimenez:2001gg}
R.~Jimenez and A.~Loeb,
``Constraining cosmological parameters based on relative galaxy ages,''
Astrophys. J. \textbf{573}, 37-42 (2002)
doi:10.1086/340549
[arXiv:astro-ph/0106145 [astro-ph]].

\bibitem{Moresco:2012}
M.~Moresco et al.,
``Improved constraints on the expansion rate of the Universe up to $z\sim1.1$ from the spectroscopic evolution of cosmic chronometers,''
JCAP, Issue 08, article id. 006 (2012)
doi:10.1088/1475-7516/2012/08/006 
[arXiv:1201.3609 [astro-ph.CO]].

\bibitem{Moresco:2015}
M.~Moresco,
``Raising the bar: new constraints on the Hubble parameter with
cosmic chronometers at $z\sim2$,''
Mon. Not. Roy. Astron. Soc. \textbf{450}, p. L16-L20 (2015)
doi:10.1093/mnrasl/slv037
[arXiv:astro-ph/0302560v1 [astro-ph.CO]].

\bibitem{Moresco:2016}
M.~Moresco et al.,
``A 6\% measurement of the Hubble parameter at $z\sim0.45$: direct evidence of the epoch of cosmic re-acceleration,''
JCAP, Issue 05, article id. 014 (2016)
doi:10.1088/1475-7516/2016/05/014 
[arXiv:1601.01701  [astro-ph.CO]].

\bibitem{Moresco:2020}
M.~Moresco,  R.~Jimenez, L.~Verde, A.~Cimatti and L.~Pozzetti,
``Setting the Stage for Cosmic Chronometers. II. Impact of Stellar Population Synthesis Models Systematics and Full Covariance Matrix,''
The Astrophy. Journal, \textbf{898}, Issue 1, id.82 (2020)
doi:10.3847/1538-4357/ab9eb0
[arXiv:2003.07362v2 [astro-ph.GA]].

\bibitem{Favale:2023lnp}
A.~Favale, A.~G\'omez-Valent and M.~Migliaccio,
``Cosmic chronometers to calibrate the ladders and measure the curvature of the Universe. A model-independent study,''
[arXiv:2301.09591 [astro-ph.CO]].

\bibitem{Brout:2022vxf}
D.~Brout, D.~Scolnic, B.~Popovic, A.~G.~Riess, J.~Zuntz, R.~Kessler, A.~Carr, T.~M.~Davis, S.~Hinton and D.~Jones, \textit{et al.}
``The Pantheon+ Analysis: Cosmological Constraints,''
Astrophys. J. \textbf{938}, no.2, 110 (2022)
doi:10.3847/1538-4357/ac8e04
[arXiv:2202.04077 [astro-ph.CO]].

\bibitem{Scolnic:2021amr}
D.~Scolnic, D.~Brout, A.~Carr, A.~G.~Riess, T.~M.~Davis, A.~Dwomoh, D.~O.~Jones, N.~Ali, P.~Charvu and R.~Chen, \textit{et al.}
``The Pantheon+ Analysis: The Full Data Set and Light-curve Release,''
Astrophys. J. \textbf{938}, no.2, 113 (2022)
doi:10.3847/1538-4357/ac8b7a
[arXiv:2112.03863 [astro-ph.CO]].

\bibitem{Scolnic:pant}
\url{https://pantheonplussh0es.github.io/}

\bibitem{Leach2006}
J.~A.~Leach, S.~Carloni and P.~K.~S.~Dunsby,
``Shear dynamics in Bianchi I cosmologies with R**n-gravity,''
Class. Quant. Grav. \textbf{23}, 4915-4937 (2006)
doi:10.1088/0264-9381/23/15/011
[arXiv:gr-qc/0603012 [gr-qc]].
  
\bibitem{Banik2016}
D.~K.~Banik, S.~K.~Banik and K.~Bhuyan,
``Dynamics of Bianchi I cosmologies in f(R) gravity in the Palatini formalism,''
Indian J. Phys. \textbf{91}, no.1, 109-119 (2017)
doi:10.1007/s12648-016-0898-6
  
\bibitem{Maartens:1994pb}
R.~Maartens and D.~R.~Taylor,
``Fluid dynamics in higher order gravity,''
Gen. Rel. Grav. \textbf{26}, 599-613 (1994)
doi:10.1007/BF02108001

\bibitem{Boisseau:2010pd}
B.~Boisseau,
``Exact cosmological solution of a Scalar-Tensor Gravity theory compatible with the $\Lambda CDM$ model,''
Phys. Rev. D \textbf{83}, 043521 (2011)
doi:10.1103/PhysRevD.83.043521
[arXiv:1011.2915 [astro-ph.CO]].

\bibitem{Akarsu:2019pvi}
\"O.~Akarsu, N.~Kat\i{}rc\i{}, N.~\"Ozdemir and J.~A.~V\'azquez,
``Anisotropic massive Brans-Dicke gravity extension of the standard $\Lambda$CDM model,''
Eur. Phys. J. C \textbf{80}, no.1, 32 (2020)
doi:10.1140/epjc/s10052-019-7580-z
[arXiv:1903.06679 [gr-qc]].

\bibitem{Schiavone:2022wvq}
T.~Schiavone, G.~Montani and F.~Bombacigno,
``f(R) gravity in the Jordan frame as a paradigm for the Hubble tension,''
Mon. Not. Roy. Astron. Soc. \textbf{522}, no.1, L72-L77 (2023)
doi:10.1093/mnrasl/slad041
[arXiv:2211.16737 [gr-qc]].

\bibitem{Grogin:2011ua}
N.~A.~Grogin, D.~D.~Kocevski, S.~M.~Faber, H.~C.~Ferguson, A.~M.~Koekemoer, A.~G.~Riess, V.~Acquaviva, D.~M.~Alexander, O.~Almaini and M.~L.~N.~Ashby, \textit{et al.}
``CANDELS: The Cosmic Assembly Near-infrared Deep Extragalactic Legacy Survey,''
Astrophys. J. Suppl. \textbf{197}, 35 (2011)
doi:10.1088/0067-0049/197/2/35
[arXiv:1105.3753 [astro-ph.CO]].

\bibitem{Barro:2019}
G.~Barro, P.~G.~Perez-Gonzalez, A.~Cava, G.~Brammer, V.~Pandya, C.~E.~Moral, P.~Esquej, H.~Dominguez-Sanchez, B.~A.~Pampliega, Y.~Guo, \textit{et al.}
``The CANDELS/SHARDS multi-wavelength catalog in GOODS-N: Photometry, Photometric Redshifts, Stellar Masses, Emission line fluxes and Star Formation Rates,''
Astrophys. J. Suppl. \textbf{243}, 22 (2019) 
doi:10.3847/1538-4365/ab23f2
[arXiv:1908.00569 [astro-ph.GA]].

\bibitem{Santini:2015}
P.~Santini, H.~C.~Ferguson, A.~Fontana, B.~Mobasher, G.~Barro, M.~Castellano, S.~L.~Finkelstein, A.~Grazian, L.~T.~Hsu, B.~Lee, \textit{et al.}
``Stellar masses from the CANDELS survey: the GOODS-South and UDS fields,''
Astrophys. J. \textbf{801}, 97 (2015) 
doi:10.1088/0004-637X/801/2/97
[arXiv:1412.5180 [astro-ph.GA]].

\bibitem{Nayyeri:2017}
H.~Nayyeri, S.~Hemmati, B.~Mobasher, H.~C.~Ferguson, A.~Cooray, G.~Barro, S.~M.~Faber, M.~Dickinson, A.~M.~Koekemoer, M.~Peth, \textit{et al.}
``CANDELS Multiwavelength Catalogs: Source Identification and Photometry in the CANDELS COSMOS Survey Field,''
Astrophys. J. Suppl. \textbf{228}, 7 (2017) 
doi:10.3847/1538-4365/228/1/7
[arXiv:1612.07364 [astro-ph.GA]].

\bibitem{Simon:2005}
J.~Simon, L.~Verde and R.~Jimenez,
``Constraints on the redshift dependence of the dark energy potential,''
Phys. Rev. D \textbf{71}, 123001 (2005)
doi:10.1103/PhysRevD.71.123001
[arXiv:astro-ph/0412269 [astro-ph]].

\bibitem{Shen:2011}
Y.~Shen, P.~B.~Hall, G.~T.~Richards, D.~P.~Schneider, M.~A.~Strauss, S.~Snedden, D.~Bizyaev, H.~Brewington, V.~Malanushenko and E.~Malanushenko, \textit{et al.}
``A Catalog of Quasar Properties from SDSS DR7,''
Astrophys. J. Suppl. \textbf{194}, 45 (2011)
doi:10.1088/0067-0049/194/2/45
[arXiv:1006.5178 [astro-ph.CO]].

\bibitem{Shen:2019}
Y.~Shen, J.~Wu, L.~Jiang, E.~Banados, X.~Fan, L.~C.~Ho, D.~A.~Riechers, M.~A.~Strauss, B.~Venemans and M.~Vestergaard, \textit{et al.}
``Gemini GNIRS Near-infrared Spectroscopy of 50 Quasars at z\ensuremath{\gtrsim}5.7,''
Astrophys. J. \textbf{873}, 35 (2019)
doi:10.3847/1538-4357/ab03d9
[arXiv:1809.05584 [astro-ph.GA]].

\bibitem{Mazzucchelli:2017}
C.~Mazzucchelli, E.~Bañados, B.~P.~Venemans, R.~Decarli, E.~P.~Farina, F.~Walter, A-C.~Eilers, H-W.~Rix, R.~Simcoe, D.~Stern, \textit{et al.}
``Physical properties of 15 quasars at z\ensuremath{\gtrsim}6.5,''
Astrophys. J. \textbf{849}, 91 (2017) 
doi:10.3847/1538-4357/aa9185
[arXiv:1710.01251 [astro-ph.GA]].

\bibitem{Banados:2017unc}
E.~Banados, B.~P.~Venemans, C.~Mazzucchelli, E.~P.~Farina, F.~Walter, F.~Wang, R.~Decarli, D.~Stern, X.~Fan and F.~Davies, \textit{et al.}
``An 800-million-solar-mass black hole in a significantly neutral Universe at redshift 7.5,''
Nature \textbf{553}, no.7689, 473-476 (2018)
doi:10.1038/nature25180
[arXiv:1712.01860 [astro-ph.GA]].

\bibitem{Matsuoka:2019a}
Y.~Matsuoka, M.~Onoue, N.~Kashikawa, M.~A.~Strauss, K.~Iwasawa, C-H.~Lee, M.~Imanishi, T.~Nagao, M.~Akiyama, N.~Asami, \textit{et al.}
``Discovery of the First Low-Luminosity Quasar at z\ensuremath{>}7,''
Astrophys. J. Lett. \textbf{872}, L2 (2019) 
doi:10.3847/2041-8213/ab0216
[arXiv:1901.10487 [astro-ph.GA]].

\bibitem{Mortlock:2011}
D.~J.~Mortlock, S.~J.~Warren, B.~P.~Venemans, M.~Patel, P.~C.~Hewett, R.~G.~McMahon, C.~Simpson, T.~Theuns, E.~A.~Gonzales-Solares and A.~Adamson, \textit{et al.}
``A luminous quasar at a redshift of z = 7.085,''
Nature \textbf{474}, 616-619 (2011)
doi:10.1038/nature10159
[arXiv:1106.6088 [astro-ph.CO]].

\bibitem{Wang:2018}
F.~Wang, J.~Yang, X.~Fan, M.~Yue, X-B.~Wu, J-T.~Schindler, F.~Bian, J-T.~Li, E.~P.~Farina, E.~Bañados, \textit{et al.}
``The Discovery of A Luminous Broad Absorption Line Quasar at A Redshift of 7.02,''
Astrophys. J. Lett. \textbf{869}, L9 (2018)
doi:10.3847/2041-8213/aaf1d2
[arXiv:1810.11925 [astro-ph.GA]].

\bibitem{Yang:2019}
J.~Yang, F.~Wang, X.~Fan, M.~Yue, X-B.~Wu, J.~Li, F.~Bian, L.~Jiang, E.~Bañados, Y.~Beletsky
``Exploring Reionization-Era Quasars IV: Discovery of Six New z\ensuremath{\gtrsim}6.5 Quasars with DES, VHS and unWISE Photometry,''
Astron. J. \textbf{157}, 236 (2019)
doi:10.3847/1538-3881/ab1be1
[arXiv:1811.11915 [astro-ph.GA]].

\bibitem{Matsuoka:2019b}
Y.~Matsuoka, K.~Iwasawa, M.~Onoue, N.~Kashikawa, M.~A.~Strauss, C-H.~Lee, M.~Imanishi, T.~Nagao, M.~Akiyama, N.~Asami, \textit{et al.}
``Subaru High-z Exploration of Low-Luminosity Quasars (SHELLQs). X. Discovery of 35 Quasars and Luminous Galaxies at 5.7 {\ensuremath{\leq}} z {\ensuremath{\leq}} 7.0,''
Astrophys. J. \textbf{883}, 183 (2019) 
doi:10.3847/1538-4357/ab3c60
[arXiv:1908.07910 [astro-ph.GA]].

\bibitem{Yang:2020}
J.~Yang, F.~Wang, X.~Fan, J.~F.~Hennawi, F.~B.~Davies, M.~Yue, E.~Banados, X-B.~Wu, B.~Venemans, A.~J.~Barth, \textit{et al.}
``Pōniuā'ena: A Luminous z=7.5 Quasar Hosting a 1.5 Billion Solar Mass Black Hole,''
Astrophys. J. Lett. \textbf{897}, L14 (2020) 
doi:10.3847/2041-8213/ab9c26
[arXiv:2006.13452 [astro-ph.GA]].

\bibitem{Wang:2021}
F.~Wang, J.~Yang, X.~Fan, J.~F.~Hennawi, A~J.~Barth, E.~Banados, F.~Bian, K.~Boutsia, T.~Connor, F.~B.~Davies, \textit{et al.}
``A Luminous Quasar at Redshift 7.642,''
Astrophys. J. Lett. \textbf{907}, L1 (2021) 
doi:10.3847/2041-8213/abd8c6
[arXiv:2101.03179 [astro-ph.GA]].

\bibitem{Zhang:2014}
C.~Zhang, H.~Zhang, S.~Yuan, S.~Liu, T.J.~Zhang and Y.C.~Sun,
``Four new observational H(z) data from luminous red galaxies in the Sloan Digital Sky Survey data release seven,''
Research in Astronomy and Astrophysics, Volume 14, Issue 10, article id. 1221-1233 (2014)
doi:10.1088/1674-4527/14/10/002
[arXiv:1207.4541v3 [astro-ph.CO]].

\bibitem{Jimenez:2003}
R.~Jimenez, L.~Verde, T.~Treu and D.~Stern,
``Constraints on the Equation of State of Dark Energy and the Hubble Constant from Stellar Ages and the Cosmic Microwave Background,''
The Astrophy. Journal, \textbf{593}, Issue 2, pp. 622-629 (2003)
doi:10.1086/376595
[arXiv:astro-ph/0302560v1 [astro-ph]].

\bibitem{Ratsimbazafy:2017}
A.L.~Ratsimbazafy et al.,
``Age-dating luminous red galaxies observed with the Southern African Large Telescope,''
Mon. Not. Roy. Astron. Soc., \textbf{467}, Issue 3, p.3239-3254 (2017)
doi:10.1093/mnras/stx301 
[arXiv:1702.00418 [astro-ph.CO]].

\bibitem{Stern:2010}
D.~Stern, R.~Jimenez, L.~Verde, M.~Kamionkowski and S.A.~Standford, 
``Cosmic chronometers: constraining the equation of state of dark energy. I: H(z) measurements,''
JCAP, Issue 02, id. 008 (2010)
doi:10.1088/1475-7516/2010/02/008 
[arXiv:0907.3149 [astro-ph.CO]].

\bibitem{Borghi:2022}
N.~Borghi, M.~Moresco and A.~Cimatti,
``Toward a Better Understanding of Cosmic Chronometers: A New Measurement of H(z) at z \ensuremath{\sim} 0.7,''
Astrophys. J. Lett. \textbf{928}, no.1, L4 (2022)
doi:10.3847/2041-8213/ac3fb2
[arXiv:2110.04304 [astro-ph.CO]].

\bibitem{Lovick:2023tnv}
T.~Lovick, S.~Dhawan and W.~Handley,
``Non-Gaussian Likelihoods for Type Ia Supernovae Cosmology: Implications for Dark Energy and $H_0$,''
[arXiv:2312.02075 [astro-ph.CO]].

\bibitem{Alonso-Lopez:2023hkx}
D.~Alonso-L\'opez, J.~de Cruz P\'erez and A.~L.~Maroto,
``A unified TDiff invariant field theory for the dark sector,''
[arXiv:2311.16836 [astro-ph.CO]].

\bibitem{Padilla:2019} 
L.E.~Padilla, L.O.~Tellez, L.A.~Escamilla and J.A.~Vazquez,
``Cosmological parameter inference with Bayesian statistics,"
Universe 2021, 7(7), 213
doi: 10.3390/Universe7070213
[arXiv:1903.11127 [astro-ph.CO]].

\bibitem{Hogg:2010} 
D.~W.~Hogg, J.~Bovy and D.~Lang,
``Data analysis recipes: Fitting a model to data,''
[arXiv:1008.4686 [astro-ph.IM]].

\bibitem{Buchner:2021}
J.~Buchner,
``UltraNest - a robust, general purpose Bayesian inference engine,''
Journal of Open Source Software, \textbf{6}, issue 60, id. 3001 (2021),
doi:10.21105/joss.03001 
[arXiv:2101.09604 [stat.CO]].

\bibitem{Buchner:2014}
J.~Buchner,
``A statistical test for Nested Sampling algorithms,''
Statistics and Computing,\textbf{26}, Issue 1-2, pp. 383-392 (2014),
doi:10.1007/s11222-014-9512-y 
[arXiv:1407.5459 [stat.CO]].

\bibitem{Buchner:2017}
J.~Buchner,
``Collaborative Nested Sampling: Big Data versus Complex Physical Models,''
Public. of the Astron. Soci. of the Pacific, \textbf{131}, Issue 1004, pp. 108005 (2019),
doi:10.1088/1538-3873/aae7fc 
[arXiv:1707.04476 [stat.CO]].

\bibitem{Mukhanov:2005sc}
V.~Mukhanov,
``Physical Foundations of Cosmology,''
Cambridge University Press, 2005,
ISBN 978-0-521-56398-7
doi:10.1017/CBO9780511790553

\bibitem{Jimenez:2019onw}
R.~Jimenez, A.~Cimatti, L.~Verde, M.~Moresco and B.~Wandelt,
``The local and distant Universe: stellar ages and $H_0$,''
JCAP \textbf{03}, 043 (2019)
doi:10.1088/1475-7516/2019/03/043
[arXiv:1902.07081 [astro-ph.CO]].

\bibitem{Valcin:2020vav}
D.~Valcin, J.~L.~Bernal, R.~Jimenez, L.~Verde and B.~D.~Wandelt,
``Inferring the Age of the Universe with Globular Clusters,''
JCAP \textbf{12}, 002 (2020)
doi:10.1088/1475-7516/2020/12/002
[arXiv:2007.06594 [astro-ph.CO]].

\bibitem{Wei:2022plg}
J.~J.~Wei and F.~Melia,
``Exploring the Hubble Tension and Spatial Curvature from the Ages of Old Astrophysical Objects,''
Astrophys. J. \textbf{928}, no.2, 165 (2022)
doi:10.3847/1538-4357/ac562c
[arXiv:2202.07865 [astro-ph.CO]].

\bibitem{Costa:2023cmu}
A.~A.~Costa, Z.~Ren and Z.~Yin,
``A bias using the ages of the oldest astrophysical objects to address the Hubble tension,''
Eur. Phys. J. C \textbf{83}, no.9, 875 (2023)
doi:10.1140/epjc/s10052-023-12038-0
[arXiv:2306.01234 [astro-ph.CO]].

\bibitem{CurtisLake:2023}
E.~Curtis-Lake et al.,
``Spectroscopic confirmation of four metal-poor galaxies at z=10.3-13.2,''
[arXiv:2212.04568v2 [astro-ph.GA]].

\bibitem{Lewis:2019xzd}
A.~Lewis,
``GetDist: a Python package for analysing Monte Carlo samples,''
[arXiv:1910.13970 [astro-ph.IM]].

\bibitem{Bromm:2009uk}
V.~Bromm, N.~Yoshida, L.~Hernquist and C.~F.~McKee,
``The formation of the first stars and galaxies,''
Nature \textbf{459}, 49-54 (2009)
doi:10.1038/nature07990
[arXiv:0905.0929 [astro-ph.CO]].

\bibitem{Rubin:2023ovl}
D.~Rubin, G.~Aldering, M.~Betoule, A.~Fruchter, X.~Huang, A.~G.~Kim, C.~Lidman, E.~Linder, S.~Perlmutter and P.~Ruiz-Lapuente, \textit{et al.}
``Union Through UNITY: Cosmology with 2,000 SNe Using a Unified Bayesian Framework,''
[arXiv:2311.12098 [astro-ph.CO]].

\bibitem{DES:2024tys}
T.~M.~C.~Abbott \textit{et al.} [DES],
``The Dark Energy Survey: Cosmology Results With \textasciitilde{}1500 New High-redshift Type Ia Supernovae Using The Full 5-year Dataset,''
[arXiv:2401.02929 [astro-ph.CO]].

\bibitem{Wang:2023gla}
J.~Wang, Z.~Huang, L.~Huang and J.~Liu,
``Quantifying the tension between cosmological models and JWST red candidate massive galaxies,''
[arXiv:2311.02866 [astro-ph.CO]].

\bibitem{Farrah:2023opk}
D.~Farrah, K.~S.~Croker, G.~Tarl\'e, V.~Faraoni, S.~Petty, J.~Afonso, N.~Fernandez, K.~A.~Nishimura, C.~Pearson and L.~Wang, \textit{et al.}
``Observational Evidence for Cosmological Coupling of Black Holes and its Implications for an Astrophysical Source of Dark Energy,''
Astrophys. J. Lett. \textbf{944}, no.2, L31 (2023)
doi:10.3847/2041-8213/acb704
[arXiv:2302.07878 [astro-ph.CO]].

\bibitem{Croker:2021duf}
K.~S.~Croker, M.~J.~Zevin, D.~Farrah, K.~A.~Nishimura and G.~Tarle,
``Cosmologically Coupled Compact Objects: A Single-parameter Model for LIGO\textendash{}Virgo Mass and Redshift Distributions,''
Astrophys. J. Lett. \textbf{921}, no.2, L22 (2021)
doi:10.3847/2041-8213/ac2fad
[arXiv:2109.08146 [gr-qc]].

\bibitem{Croker:2020plg}
K.~S.~Croker, J.~Runburg and D.~Farrah,
``Implications of Symmetry and Pressure in Friedmann Cosmology. III. Point Sources of Dark Energy that Tend toward Uniformity,''
Astrophys. J. \textbf{900}, no.1, 57 (2020)
doi:10.3847/1538-4357/abad2f

\bibitem{Gliner:1966}
E.~B.~Gliner,
``Algebraic Properties of the Energy-Momentum Tensor and Vacuum-like States of Matter,''
Sov.Phys.JETP 22 (1966) 378.

\bibitem{Nunes:2020hzy}
R.~C.~Nunes, S.~K.~Yadav, J.~F.~Jesus and A.~Bernui,
``Cosmological parameter analyses using transversal BAO data,''
Mon. Not. Roy. Astron. Soc. \textbf{497}, no.2, 2133-2141 (2020)
doi:10.1093/mnras/staa2036
[arXiv:2002.09293 [astro-ph.CO]].

\bibitem{eBOSS:2020yzd}
S.~Alam \textit{et al.} [eBOSS],
``Completed SDSS-IV extended Baryon Oscillation Spectroscopic Survey: Cosmological implications from two decades of spectroscopic surveys at the Apache Point Observatory,''
Phys. Rev. D \textbf{103}, no.8, 083533 (2021)
doi:10.1103/PhysRevD.103.083533
[arXiv:2007.08991 [astro-ph.CO]].

\bibitem{deCarvalho:2021azj}
E.~de Carvalho, A.~Bernui, F.~Avila, C.~P.~Novaes and J.~P.~Nogueira-Cavalcante,
``BAO angular scale at zeff = 0.11 with the SDSS blue galaxies,''
Astron. Astrophys. \textbf{649}, A20 (2021)
doi:10.1051/0004-6361/202039936
[arXiv:2103.14121 [astro-ph.CO]].

\bibitem{DiValentino:2020vvd}
E.~Di Valentino, L.~A.~Anchordoqui, \"O.~Akarsu, Y.~Ali-Haimoud, L.~Amendola, N.~Arendse, M.~Asgari, M.~Ballardini, S.~Basilakos and E.~Battistelli, \textit{et al.}
``Cosmology Intertwined III: $f \sigma_8$ and $S_8$,''
Astropart. Phys. \textbf{131}, 102604 (2021)
doi:10.1016/j.astropartphys.2021.102604
[arXiv:2008.11285 [astro-ph.CO]].

\bibitem{Nunes:2021ipq}
R.~C.~Nunes and S.~Vagnozzi,
``Arbitrating the S8 discrepancy with growth rate measurements from redshift-space distortions,''
Mon. Not. Roy. Astron. Soc. \textbf{505}, no.4, 5427-5437 (2021)
doi:10.1093/mnras/stab1613
[arXiv:2106.01208 [astro-ph.CO]].

\bibitem{Esposito:2022plo}
M.~Esposito, V.~Ir\v{s}i\v{c}, M.~Costanzi, S.~Borgani, A.~Saro and M.~Viel,
``Weighing cosmic structures with clusters of galaxies and the intergalactic medium,''
Mon. Not. Roy. Astron. Soc. \textbf{515}, no.1, 857-870 (2022)
doi:10.1093/mnras/stac1825
[arXiv:2202.00974 [astro-ph.CO]].

\bibitem{Adil:2023jtu}
S.~A.~Adil, \"O.~Akarsu, M.~Malekjani, E.~\'O.~Colg\'ain, S.~Pourojaghi, A.~A.~Sen and M.~M.~Sheikh-Jabbari,
``$S_8$ increases with effective redshift in $\Lambda$CDM cosmology,''
doi:10.1093/mnrasl/slad165
[arXiv:2303.06928 [astro-ph.CO]].

\end{thebibliography}
\end{document}